%% file: AlahmedTongZhao_v8.tex
\documentclass[journal]{IEEEtran}
\usepackage[left=0.75in, right=0.75in, top=1in, bottom=0.9in]{geometry}
\makeatletter
\def\ps@headings{%
\def\@oddhead{\mbox{}\scriptsize\rightmark \hfil \thepage}%
\def\@evenhead{\scriptsize\thepage \hfil \leftmark\mbox{}}%
\def\@oddfoot{}%
\def\@evenfoot{}}
\makeatother 

\usepackage{amsmath,amssymb,mathrsfs,amsthm,bm}
\usepackage{mathtools}
\usepackage{cite}

\usepackage{graphicx,stfloats,subfigure,multicol}
\usepackage{epsfig,epsf,psfrag,amssymb,amsfonts,latexsym,graphicx,mathrsfs,subfigure}

\usepackage{comment}
\usepackage{lettrine}
\usepackage{enumerate}
\usepackage{booktabs}
\usepackage[table]{xcolor}
\usepackage{multirow}
\usepackage{tablefootnote}
\usepackage{bbm}
\usepackage{chngpage}
\usepackage{textcomp}
\usepackage{hyperref}
\usepackage{orcidlink}
\usepackage{algorithm,algpseudocode}
\usepackage{url}
\usepackage[hyphenbreaks]{breakurl}
\usepackage[normalem]{ulem}
\usepackage{algpseudocode}
\newsavebox{\ieeealgbox}

\usepackage{array}

\newtheorem{theorem}{Theorem}
\newtheorem{proposition}{Proposition}
\newtheorem{corollary}{Corollary}
\newtheorem{lemma}{Lemma}

\newcommand*{\QED}{\hfill\ensuremath{\square}}

 \def\old#1{}

\input LTmacros

\begin{document}

\title{Co-Optimizing Distributed Energy Resources in Linear Complexity under Net Energy Metering}

\author{Ahmed S. Alahmed\orcidlink{0000-0002-4715-4379},~\IEEEmembership{Student~Member,~IEEE},
Lang~Tong\orcidlink{0000-0003-3322-2681},~\IEEEmembership{Fellow,~IEEE},
Qing~Zhao\orcidlink{0000-0002-9590-4285},~\IEEEmembership{Fellow,~IEEE}
\thanks{\scriptsize  Ahmed S. Alahmed,
Lang Tong, and Qing Zhao  ({\tt \{\tcb{ASA278,~LT35,~QZ16}\}\tcb{@cornell.edu}}) are  with the School of Electrical and Computer Engineering, Cornell University, USA.
This work was supported in part by the National Science Foundation under Grant 2218110.}
\vspace{-0.45cm}
}
\maketitle

{
\begin{abstract}
The co-optimization of behind-the-meter distributed energy resources is considered for prosumers under the net energy metering tariff. The distributed energy resources considered include renewable generations, flexible demands, and battery energy storage systems. An energy management system co-optimizes the consumptions and battery storage based on locally available stochastic renewables by solving a stochastic dynamic program that maximizes the expected operation surplus. To circumvent the exponential complexity of the dynamic program solution, we propose a closed-form and linear computation complexity co-optimization algorithm based on a relaxation-projection approach to a constrained stochastic dynamic program.  Sufficient conditions for optimality for the proposed solution are obtained. Numerical studies demonstrate orders of magnitude reduction of computation costs and significantly reduced optimality gap.
\end{abstract}
\begin{IEEEkeywords}
Battery storage systems, distributed energy resources, dynamic programming, energy management systems, flexible demands, Markov decision process, net energy metering.
\end{IEEEkeywords}
}

\section{Introduction} \label{sec:intro}
\input intro_v7

\section{Co-optimization Problem Formulation} \label{sec:formulation}
\input formulation_v7

\section{Linear-Complexity Myopic Co-optimization}\label{sec:myopic}
\input Myopic_v7.tex

\section{Optimal Solution Structure and Insights} \label{sec:prosumer}
\input prosumer_v7

\section{Numerical Results}\label{sec:num}
\input num_v7


\section{Conclusion}\label{sec:conclusion}

This paper presents the first decentralized linear complexity solution to the DER co-optimization of BTM storage and consumption decisions under the general NEM-X tariff models. The developed technique applies more broadly to the residential and commercial/industrial energy management systems and is optimal
under certain operating conditions. The results demonstrate that the widely deployed NEM tariff leads to rich structural properties of the co-optimization and results in a highly effective, low-cost, and scalable solution to an otherwise intractable stochastic dynamic programming problem. Although not always optimal, the developed approach shows promising performance in simulations. 

Several theoretical and practical issues remain, showing some limitations of the proposed approach. One limitation is that the proposed model excludes the optimization of certain deferrable loads, such as EV charging with deadlines. Such problems cannot be modeled directly under the optimization problem presented here; a non-trivial modification is needed, as shown in \cite{Jeon&Tong&Zhao:23PESGM}. Another limitation is that it is not obvious how the practical implementation of the algorithm can accommodate flexible loads with non-linear operating conditions.
Lastly, the impact of DER co-optimization benefits not only the prosumer but also the grid operator and utilities in the form of the reduction of RPF and utility's net costs (see Appendix \ref{App:ProofsC}). Expanded characterization of such social benefits requires a separate study.

\appendices

\section{Nomenclature} \label{sec:nomenclature}
\input appendix_Nomenclature

\section{Preliminaries and Proofs} \label{App:ProofsA}
\input appendixA_v7

\section{Optimal Prosumer Decisions under Relaxation of (\ref{eq:gamma})} \label{App:ProofsB}
\input appendixB_v7

\section{Incorporating Storage Degradation} \label{App:Degrade}
\input appendixD_v7

\section{Benefits of Co-optimization} \label{App:ProofsC}
\input appendixC_v7

{
\bibliographystyle{IEEEtran}
\bibliography{Storage_Tong}
}

\end{document}

%% file: LTmacros.tex
\def\nn{\nonumber}
\def\beq{\begin{equation}}
\def\eeq{\end{equation}}
\def\bea{\begin{eqnarray}}
\def\eea{\end{eqnarray}}
\def\ba{\begin{array}}
\def\ea{\end{array}}

\def\bitem{\begin{itemize}}
\def\eitem{\end{itemize}}
\def\ben{\begin{enumerate}}
\def\een{\end{enumerate}}

\def\ie{{\it i.e.,\ \/}}

\def\tcb{\textcolor{blue}}

\newcommand{\mbbE}{\mathbb{E}}

\newcommand{\mbbR}{\mathbb{R}}

\newcommand{\Emsc}{\mathscr{E}}

\def\Ac{{\cal A}}

\def\Gc{{\cal G}}

\def\Lc{{\cal L}}

\def\Oc{{\cal O}}
\def\Pc{{\cal P}}

%% file: intro_v7.tex
\lettrine{E}{nergy} sustainability entails full coordination and utilization of the flexible behind-the-meter (BTM) distributed energy resources (DER) \cite{Asimakopoulou&Hatziargyriou:18TSE,Khezri&Mahmoudi&Haque:21TSE}. In this work, we consider the problem of co-optimizing BTM DER that include flexible demand, renewable distributed generation (DG), and BTM storage motivated by the increasing electrification in distribution networks with flexible consumption \cite{LoadFlexibility:23ESIG}, battery systems \cite{vandeVen&etal:13TSG} and the potential of aggregated DER in achieving energy sustainability and resiliency \cite{Asimakopoulou&Hatziargyriou:18TSE,FERC2222}.

\begin{figure}
    \centering
    \includegraphics[scale=0.45]{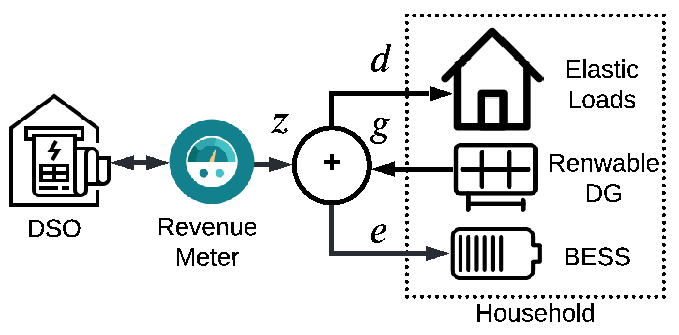}
    \vspace{-0.34cm}
    \caption{Net energy metering scheme, with $d, g \in \mathbb{R}_+$ being the variables of consumption and renewable DG, respectively, whereas $e, z \in \mathbb{R}_+$ are the variables of storage output and net consumption, respectively.  Arrows of variables correspond to the positive direction.}
    \label{fig:MeteringArchitecture}
    \vspace{-0.45cm}
\end{figure}

The settings we have in mind are smart homes with DER as illustrated in Fig.~\ref{fig:MeteringArchitecture}, where an intelligent energy management system (EMS) optimizes the prosumer's consumption bundle and storage operation, given available renewables, consumption preferences and flexibility, storage capabilities, and the retail price of electricity from a distribution system operator (DSO) or an aggregator. In particular, we are interested in the net energy metering (NEM) retail program that determines the payment based on the net energy consumption measured by the revenue meter. A review of NEM-based tariffs and their impacts on prosumer decisions can be found in \cite{NEMevolution:23NAS, AlahmedTong:22EIRACM}.

Our setting also includes EMS for buildings and small campuses of commercial or industrial customers, including community solar/storage programs.  Also relevant is the DER aggregation for participation in the wholesale electricity market as articulated in the Federal Energy Regulatory Commission (FERC) order 2222 \cite{FERC2222}.  In that context, the technique developed here can be used by DER aggregators or energy communities to schedule a large number of small but ubiquitous DGs and flexible demands \cite{Alahmed&Tong:24TEMPR}.

Co-optimizing BTM DER becomes more imperative as the difference between the export and import prices under NEM becomes larger. This is because the value of self-consuming, or storing, the renewable DG becomes more valuable compared to exporting it to the grid. The challenge of co-optimizing BTM DER is twofold.  First,  the optimal real-time scheduling of consumption and storage operations requires knowing the underlying probabilistic models of the renewables.  Despite recent advances in machine learning technologies, learning the optimal scheduling policy from historical data or online reinforcement learning remains difficult.  Existing techniques typically do not have performance guarantees appropriate for practical applications, nor reasonable computation and sample complexities of learning \cite{Peddrasa&etal:10TSG, Grijalva&Tanguy:12IEEETSG}.

Second, scalable solutions are necessary for co-optimizing a large number of controllable energy devices in building
or microgrid-level co-optimizations. The computation and coordination costs may be prohibitive in practice. To this end, linear complexity co-optimization algorithms with possibly decentralized low-cost hardware implementation and minimal communication overhead are highly desirable. To our best knowledge, no such solutions currently exist.

\subsection{Related Work}
 There is significant literature on managing BTM energy storage \cite{Harshah&Dahleh:15TPS,Kwon&Xu&Gautam:17TSG,vandeVen&etal:13TSG,Chen&Wang&Heo&Kishore:13TSG}, and managing flexible demands (i.e., demand side management) \cite{Jeddi&Mishra&Ledwich:20TSE,Alahmed&Tong:22TSG,Khezri&Mahmoudi&Haque:21TSE} when co-existing with other DERs, although the literature on co-optimizing BTM DERs is relatively limited, with no prior analytical work in the context of the retail electricity market under NEM tariffs and no existing co-optimization algorithms with linear computation costs.  Here, we discuss some of the existing approaches having similar optimization structures as that considered in this work.
 


We classify existing DER optimization solutions into static and dynamic optimizations.  Under static optimizations, one of the earliest storage and flexible demand co-optimization formulations is \cite{Peddrasa&etal:10TSG}, where the authors optimize across several types of BTM DER, including rooftop solar, thermal storage, and flexible demands to maximize customer surplus.  The optimization is deterministic, assuming perfect prediction of available rooftop solar within the scheduling horizon. The work in \cite{Grijalva&Tanguy:12IEEETSG} formulates a mixed-integer linear program that co-optimizes energy storage and flexible loads using a divide-and-conquer algorithm. In \cite{Li&Dong:19TSG}, a BTM storage management algorithm with a closed-form policy is proposed, which takes into account charging-discharging inefficiencies and effects on battery degradation. The authors, however, do not consider the flexibility of consumption and assume a sub-optimal policy that requires renewables to feed the loads first. In \cite{Li&Chen&Low:11PESGM}, storage-consumption co-optimization is considered without BTM renewables. The optimization is a deterministic linear program with a decentralized implementation.  A stochastic co-optimization of storage and renewables is formulated in \cite{Gonzalez&etal:08TPS}.  Such a formulation can be considered as a storage-consumption co-optimization if the renewables are treated as a negative demand subject to curtailment.  Proposed for a day-ahead market operation, the technique in \cite{Gonzalez&etal:08TPS} does not consider real-time scheduling based on available renewables.

The second category is stochastic dynamic programming, where the available renewables at each time $t$ are used in scheduling decisions.  Optimizing storage operations has been studied in many settings without storage-consumption co-optimization.  In \cite{vandeVen&etal:13TSG,Harshah&Dahleh:15TPS,Kwon&Xu&Gautam:17TSG}, the problem of storage management for stochastic demand is formulated as an infinite-horizon average/discounted-cost dynamic programs where renewable generation, inelastic demand, and energy prices are modeled as exogenous random processes.  
The objective is to minimize the costs of energy purchases and storage operations. The resulting solutions have the characteristics of optimal inventory control with a two-threshold policy---the so-called $(s,S)$ policy---defined on the space of the battery state-of-charge (SoC).

 Closer to the formulation considered in this work is the surplus maximization dynamic programming in \cite{Xu&Tong:17TAC} that allows prosumers to export power back to the grid under differentiated buying and selling prices. Among dynamic programming solutions, \cite{Xu&Tong:17TAC} and \cite{Guo&etal:13TSG} stand out for the co-optimization of storage and the time-of-service of the deferrable load.  We note that co-optimizing storage and time-of-service of deferrable load is very different from the co-optimization in our work, which optimizes the quantity (not the time of service) of the elastic demand. A recent work along this line under NEM tariff is \cite{Jeon&Tong&Zhao:23PESGM}.

Finally, a practical and widely-used alternative to dynamic programming is the model predictive control (MPC) proposed in \cite{Le&etal:12JEE,Chen&Wang&Heo&Kishore:13TSG,Seal&etal:23TSE, Yousefi&Hajizadeh&Soltani&Hredzak:21TII,Karthikeyan&Pillai&Jensen&Simpson:19TPS}. One of the earliest presentations is \cite{Le&etal:12JEE} where a quadratic program is solved in each interval based on the look-ahead forecast of renewables with the co-optimized storage and generation decisions in the immediate interval implemented. An MPC-based method is proposed in \cite{Chen&Wang&Heo&Kishore:13TSG} for residential appliance and storage scheduling. However, renewable DG generation and selling back to the grid feature were not considered. In \cite{Seal&etal:23TSE}, an MPC-based EMS that co-optimizes household temperature and electric vehicle and home batteries under the existence of renewable DG with the objective of minimizing household energy payment under {\em NEM 1.0} (i.e., equivalent grid buy and sell rates \cite{AlahmedTong:22EIRACM}).  Using the same objective in \cite{Seal&etal:23TSE}, the authors in \cite{Yousefi&Hajizadeh&Soltani&Hredzak:21TII} use MPC to co-optimize heat pumps and electric vehicle charging while meeting the household's comfort requirements. Lastly, heat pumps and battery energy storage are co-optimized in \cite{Karthikeyan&Pillai&Jensen&Simpson:19TPS} to provide demand response in distribution networks under the influence of network constraints. A receding horizon MPC-based method was used to control the flexible resources.

We omit the literature on price arbitrage in the wholesale/retail markets.   Under well-designed NEM tariffs,  price arbitrage is infeasible because sell (export) prices are uniformly lower than retail (consumption) prices.









Our work is built upon \cite{Alahmed&Tong:22TSG}, where the optimal consumption decision without BTM storage is considered under the general form of the NEM X tariff.  It is shown in \cite{Alahmed&Tong:22TSG} that the optimal consumption decision is a {\em two-threshold policy} based on the available BTM renewables; when the level of renewable is below the lower threshold
or above the higher threshold, the optimal {\em consumptions} are constants, and the optimal {\em net consumption} (consumption $-$ renewables) decreases piecewise linearly with the available renewables.
Between the two renewable-independent thresholds, there is a net-zero consumption zone, where the total consumption matches the BTM renewables, and the prosumer is effectively off the grid. This general characteristic carries over to the storage-consumption co-optimization, where the flexibility of both demands and energy storage is utilized to maximize prosumer's surplus.  See Sec. \ref{sec:myopic}--\ref{sec:prosumer}.

\subsection{Summary of Results and Contributions}
The main contribution of this work is a linear-complexity heuristic solution to the storage and flexible demand co-optimization, which makes it possible to schedule a large number of flexible demands and storage resources in a decentralized fashion as closed-form functions of the BTM renewables.  By linear complexity, we mean that the computation and communication costs scale linearly with the number of energy-consuming devices (the size of the consumption bundle) and the scheduling horizon.

To circumvent the intractability of solving a stochastic dynamic program that requires backward induction to co-optimize storage and flexible demands, we propose a suboptimal but linear complexity co-optimization algorithm, referred to as myopic co-optimization (MCO), that relaxes the storage SoC limits constraints and projects the problem into the feasible solution set if the solution generates an infeasible action. In particular, the MCO clips the storage charging/discharging amount to the upper or lower limits (Theorem \ref{thm:Myopic}). Significantly, the MCO does not require the underlying probability distributions of the renewable generation. We demonstrate in simulations that MCO is optimal in some cases even when SoC constraints are binding, and the performance loss is quite small.

We then propose a sufficient condition for the MCO to be optimal and obtain structural properties of the optimal co-optimization policy (Theorem \ref{thm:sufficient}). We show that under the sufficient optimality condition, the optimal policy is based on a set of global thresholds on the BTM renewables. These thresholds can be computed in closed form.  In such cases, the optimal storage operation and consumption levels of all flexible demands are expressed in closed-form as functions of the BTM renewables.

We show several novel insights of storage-demand co-optimization under NEM.  One is complementarity condition (Lemma 2), showing that, under NEM, storage charging can only happen when the prosumer is not a {\em net-consumer}, and storage discharging can only happen when the prosumer is not a {\em net-producer}.  The second is the role of the co-optimization in enlarging the net-zero zone where the prosumer is operationally ``off the grid'', which benefits DSOs in terms of reducing congestion and reverse power flows (RPF). Crucial to regulators, the characterized optimal prosumers response elucidates how the design of NEM tariff influences their consumption and storage operation decisions.

In Sec. \ref{sec:num}, we present simulation results comparing the performance of the proposed MCO algorithm with the standard MPC algorithm and several solutions offered in Tesla's Powerwall modules.  We demonstrate that significant gain in performance over the MPC benchmark because the proposed technique eliminates the need for renewable predictions.  Substantial gains were also observed over commercial solutions.


\subsection{Mathematical Notations and Paper Organization}
The paper's nomenclature is given in the appendix. Throughout the paper, we use boldface letters to indicate column vectors as in $\bm{x}=(x_1,\cdots, x_n)$. In particular, $\bm{1}$ is a column vector of all ones. For a vector $\bm{x}$,  $\bm{x}^\top$ is the transpose of $\bm{x}$.  For a multivariate function $f$ of $\bm{x}$, we use interchangeably $f(\bm{x})$ and $f(x_1,\cdots,x_n)$. We use $(x_t)$ to denote a time series indexed by $t$.  A finite sequence $(x_t)$ may be represented by a column vector denoted by $\bm{x}=(x_t)$.
For vectors $\bm{x},\bm{y}$,  $\bm{x} \preceq \bm{y}$ is the element-wise inequality $x_i \le y_i$ for all $i$, and $[\bm{x}]^+, [\bm{x}]^-$ are the element-wise positive and negative parts of vector $\bm{x}$, \ie $[x_i]^+=\max\{0,x_i\}$, $[x_i]^- =-\min\{0,x_i\}$ for all $i$, and $\bm{x}= [\bm{x}]^+ - [\bm{x}]^-$. Finally, the conditional distribution of $x_{t+1}$ given $x_t$ is expressed as $F_{x_{t+1}|x_t}$.

The rest of the paper is organized as follows. Sec.\ref{sec:formulation} presents the DER models and the DER co-optimization problem. Sec.\ref{sec:myopic} proposes the linear-complexity co-optimization algorithm and establishes the theoretical results. Sec.\ref{sec:prosumer} is used to deliver insights and intuition on the co-optimization solution structure, followed by the simulation results in Sec.\ref{sec:num} and the paper's conclusion in Sec.\ref{sec:conclusion}.

%% file: formulation_v7.tex
\subsection{BTM DER Models}
For the system model in Fig.\ref{fig:MeteringArchitecture}, we consider the sequential co-scheduling of consumption $(\bm{d}_t)$ and storage operation $(e_t)$  over a finite horizon indexed by $t=0,\ldots, T-1$. The proposed approach applies to EMS optimized locally for individual customers, hence network effects are not relevant.

\paragraph{Renewable}  The BTM  renewable generation  $(g_t)$  is an exogenous  (positive) Markovian random process. 

\paragraph{Battery storage}   The battery SoC of the  storage is denoted by $s_t \in [0,B]$ with $B$ as the maximum operational SoC. The storage control in interval $t$ is denoted by $e_t \in [-\underline{e},\bar{e}]$, where
 $\underline{e}$ and $\bar{e}$ are the maximum energy discharging and charging limits, respectively.  The battery is charged when $e_t >0$ and discharged when $e_t<0$.

The evolution of $s_t$ driven by control $e_t$ is given by
\begin{equation}\label{eq:SOCevolution}
    s_{t+1} = s_t + \tau [e_t]^+-[e_t]^-/\rho, \quad t=0,\ldots, T-1,
\end{equation}
where $\tau \in (0,1]$ and $\rho \in (0,1]$ are the charging and discharging efficiencies, respectively. For brevity, we relegate incorporating the long-term storage degradation caused by short-term charging/discharging actions to Appendix~\ref{App:Degrade}, which is done by exploiting a simple modification to the objective function in (\ref{eq:optimization}) allowing us to incorporate long-term per-unit degradation cost from charging/discharging actions.



\paragraph{Flexible demand and utility of consumption}  We assume the prosumer has $K$ controllable devices whose energy consumption bundle in interval $t$ is denoted by
\[ \bm{d}_t = (d_{t1},\cdots, d_{tK}) \in \mathcal{D}:=\{\bm{d}:\bm{0} \preceq \bm{d} \preceq \overline{\bm{d}}\} \subseteq \mathbb{R}^K_+,
\]  where $\overline{\bm{d}}$ is the consumption bundle's upper limit. The prosumer's {\em total consumption} $d_t$ and {\em net consumption} $z_t$ in interval $t$ are defined, respectively, by
\begin{equation}\label{eq:NetConsumption}
d_t:= \bm{1}^\top \bm{d}_t,~~   z_t:=d_t + e_t-g_t.
\end{equation}

Following the standard microeconomics theory, we assume the prosumer preference on consumption set $\mathcal{D}$ is characterized by a utility $U_t(\bm{d}_t)$ of consuming $\bm{d}_t$ in interval $t$ \cite{Varian2microeconomic:1992book}. $U_t(\bm{d}_t)$ is assumed to be additive, with marginal utility function denoted by $\bm{L}_t$. Specifically, for $t=0,\ldots, T-1$,
\beq 
U_t(\bm{d}_t) := \sum_{k=1}^K U_{tk}(d_{tk}),\bm{L}_t:=\nabla U_t=(L_{t1},\cdots, L_{tK}).\nn
\eeq
We assume $U_t(\bm{d}_t)$ is concave, non-decreasing, and continuously differentiable. In practice, $U_t(\bm{d}_t)$ is unknown.  Here we assume that $U_t(\bm{d}_t)$ can be learned using a variety of machine-learning algorithms. See, e.g.,  \cite{Keshavarz&Wang&Boyd:11ISIC}. 

\subsection{NEM X Tariff Model}
NEM tariff is a billing mechanism that determines prosumer credits and payments within each billing period based on the {\em net consumption} computed over the {\em net billing period}.  The net billing period can be as short as 5 minutes and as long as a day or a month.  For ease of presentation, we restrict ourselves to the case that the NEM billing period is the same as the prosumer's scheduling period, which allows us to index the billing period also by $t$.

 We adopt the NEM X tariff model proposed in \cite{Alahmed&Tong:22TSG,AlahmedTong:22EIRACM}. For $t=0,\ldots, T-1$, given the NEM X tariff parameter $\pi_t=(\pi^+_t,\pi^-_t)$, the customer's payment under NEM X is
\begin{equation}\label{eq:NEMpayment}
    P_{\pi_t}^{\mbox{\tiny NEM}}(z_t) := \pi^{+}_t [z_t]^+-\pi^{-}_t [z_t]^-,
\end{equation}
where $\pi^+_t \ge0$ is the {\em retail rate},  and $\pi^-_t \ge0$  the {\em export (compensation) rate}.
A  prosumer is a \textit{net-consumer} facing $\pi^+_t$ when $z_t \ge 0$ and a \textit{net-producer} facing $\pi_t^-$ when $z_t < 0$. Throughout this work, we assume $\max\{(\pi_t^-)\}  < \min\{(\pi_t^+)\}$, which avoids risk-free price arbitrage, as the NEM X rates are deterministic and known {\em apriori}. 


\subsection{Consumption and Storage Co-optimization}

We formulate the prosumer decision problem as a $T$-stage Markov decision process (MDP).  The state $x_t :=(s_t, g_t) \in \mathcal{X}$ of the MDP in interval $t$ includes the battery SoC $s_t$ and renewable generation $g_t$, whose evolution is defined by (\ref{eq:SOCevolution}) and the exogenous Markov random process $(g_t)$. The initial state is denoted by $x_0=(s, g)$. The randomness in this MDP formulation arises from the stochastic renewable generation.

\par An MDP {\em policy} $\mu := (\mu_0,\ldots,\mu_{T-1})$ is a sequence of decision rules, $x_t \stackrel{\mu_t}{\rightarrow} u_t := (\bm{d}_t,e_t)$, specifying consumption and storage operation in each $t$. For $t=0,\ldots, T-1$, the control action $u_t$ generates a {\em prosumer surplus} defined by
\beq
S^{\mbox{\tiny NEM}}_{\pi_t}(u_t;g_t) := U_t(\bm{d}_t)-P_{\pi_t}^{\mbox{\tiny NEM}}(\bm{1}^\top \bm{d}_t+e_t-g_t). \label{eq:surplus}
\eeq
The stage reward is defined by the prosumer surplus and the terminal salvage value of the storage:
\beq
        r_{t}\left(x_t,u_t\right) := \begin{cases} S_{\pi_t}^{\mbox{\tiny NEM}}(u_t;g_t), & t \in[0, T-1] \\ \gamma (s_{T}-s), & t=T,\end{cases} \label{eq:stageReward}
\eeq
with $\gamma$ being the marginal value of energy in the storage that represents the value of the stored energy to be used in future scheduling. With little loss of generality, we assume
\beq \label{eq:gamma}
\frac{1}{\tau} \max\{(\pi_t^-)\} \le \gamma \le \rho \min\{(\pi_t^+)\}.
\eeq
When (\ref{eq:gamma}) is not satisfied, the co-optimization policy is relatively easier to obtain and trivial in some cases. See Appendix~\ref{App:ProofsB} for a complete solution when (\ref{eq:gamma}) is relaxed, and for a discussion on why meeting (\ref{eq:gamma}) is more practical.

 The storage-consumption co-optimization is defined by
 \begin{subequations}\label{eq:optimization}
\begin{align}
   \Pc: \underset{\mu = (\mu_0,\ldots,\mu_{T-1})}{\text { Maximize}} & \mathbb{E}_{\mu}\left\{\gamma (s_{T}-s)+\sum_{t=0}^{T-1}   r_{t}\left(x_t,u_t\right)\right\} \\\text { Subject to~~} & \mbox{for all}~ t=0,\ldots, T-1,\nn\\&
   z_t = \bm{1}^\top \bm{d}_t+ [e_t]^+-[e_t]^- -g_t\\&
   s_{t+1}=s_t + \tau [e_t]^+-[e_t]^-/\rho \\&
    g_{t+1} \sim F_{g_{t+1}|g_t} \\&
    0 \leq s_{t} \leq B \label{eq:SoCcapacity}\\&
    0 \leq [e_t]^- \leq \underline{e} \label{eq:e-}\\&
    0 \leq [e_t]^+ \leq \overline{e}   \label{eq:e+}\\&
    \bm{0} \preceq \bm{d}_{t} \preceq \bm{\overline{d}} \label{subeq:DeviceConstraint}   \\&
    x_{0}=(s,g) , \end{align}
\end{subequations}
 where the expectation is taken over the exogenous stochastic generation $(g_t)$. Note that, by definition, $[e_t]^+ \cdot [e_t]^-=0$. 
 
 The {\em value function} at time $t$, denoted by $V_t(\cdot)$, maps state $x_t$ to the maximum total cumulative reward under the {\em optimal policy} from $t$ to the end of the scheduling.

We note that no linear complexity solution exists for (\ref{eq:optimization}). The optimal solution obtained from the standard backward induction has exponential complexity with respect to the state of the dynamic program, which highlights the importance of a low-complexity solution as presented in the next section. A similar stochastic dynamic program was formulated in \cite{Li&Dong:19TSG} but with a non-flexible consumption, and a sub-optimal policy that requires the renewables to feed the loads first. The work in \cite{Xu&Tong:17TAC} also considers a similar program, which, unlike \cite{Li&Dong:19TSG}, considers co-optimizing storage and consumption. However, the solution in \cite{Li&Dong:19TSG} is not closed-form and the consumption was co-optimized in terms of time-of-service of the deferrable load rather than quantity of the flexible load. 

Alternative to dynamic programming formulation, MPC-based models were proposed in  \cite{Le&etal:12JEE,Chen&Wang&Heo&Kishore:13TSG,Seal&etal:23TSE, Yousefi&Hajizadeh&Soltani&Hredzak:21TII,Karthikeyan&Pillai&Jensen&Simpson:19TPS}, mainly requiring renewable forecasts, the quality of which affects the MPC algorithm performance and computation complexity.

%% file: Myopic_v7.tex
 We propose here a linear-complexity co-optimization algorithm to solve a myopic version of the stochastic dynamic program in (\ref{eq:optimization}), hence removing the requirement to look-ahead of available renewables in the future, and the non-linear complexity of the typical $(s,S)$ optimal inventory control solutions with lower/upper thresholds $s/S$ on the SoC \cite{Sethi&Cheng:97OR}. 

Conceptually, the MCO algorithm relaxes the SoC limits constraint (\ref{eq:SoCcapacity}) that introduces strong temporal coupling, and then it projects the relaxed problem to the feasible solution set when the storage charging/discharging action
in each interval violate (\ref{eq:SoCcapacity}). In simpler words, for every $t\in [0,T-1]$, the MCO algorithm considers a natural heuristic that ``clips'' the storage charging/discharging amount when the updated storage SoC exceeds the SoC limits in (\ref{eq:SoCcapacity}), and therefore solves the following myopic program:

  \begin{subequations}\label{eq:Opt_myopic}
\begin{align}
   \Pc_t:\hspace{-0.0cm} \underset{\bm{d}_t \in \mathbb{R}^K_+,e_t\in \mathbb{R}}{\text { Maximize}} &~~ U_t(\bm{d}_t) - P_{\pi_t}^{\mbox{\tiny NEM}}(z_t) + \gamma (\tau [e_t]^+-[e_t]^-/\rho)\\
   \text { Subject to:~~} & z_t = \bm{1}^\top \bm{d}_t+ [e_t]^+-[e_t]^- -g_t \\
    & 0 \leq [e_t]^- \leq \min\{\underline{e},\rho s_t\} \label{eq:Clip1} \\
    & 0 \leq [e_t]^+ \leq \min\{\overline{e},(B-s_t)/\tau\} \label{eq:Clip2}\\
    & \bm{0} \preceq \bm{d}_{t} \preceq \bm{\bar{d}} ,
\end{align}
\end{subequations}
 where the last term of the objective function captures the terminal benefit of the energy stored in the storage, and the clipping is tackled by (\ref{eq:Clip1})-(\ref{eq:Clip2}).
 
 \subsection{Myopic Co-optimization Optimal Scheduling}
 The following theorem delineates the closed-form solution of the optimal storage and demand schedules of (\ref{eq:Opt_myopic}). The optimal schedules are all threshold policies based on the renewable $g_t$ with thresholds that can be computed {\em apriori}.
 
\begin{theorem}[Myopic co-optimization optimal scheduling]\label{thm:Myopic}
For every $t\in [0,T-1]$, the optimal storage $e^\dagger_t$ and total demand $d^\dagger_t$ under the myopic co-optimization $\Pc_t$ that is a relaxation of $\Pc$ in (\ref{eq:optimization}) are monotonically increasing piecewise-linear functions of $g_t$ with the following ordered thresholds
\begin{equation*}
\Delta_t^{+}\le \sigma_t^{+}\le \sigma_t^{+o}\le \sigma_t^{-o}\le \sigma_t^{-}\le \Delta_t^{-},
\end{equation*}
defined by
\begin{align}\label{eq:DeltaSigma}
\Delta_t^{+}&:= \max\{f_t(\pi_t^+)-\underline{e}^\dagger_t,0\},~\Delta_t^{-}:= f_t(\pi_t^-) + \overline{e}^\dagger_t,\nn\\
\sigma_t^{+}&:= \max\{f_t(\gamma/\rho)- \underline{e}^\dagger_t,0\}, \sigma_t^{-}:= f_t(\tau \gamma) + \overline{e}^\dagger_t,\\
\sigma_t^{+o}&:= f_t(\gamma/\rho),\hspace{2cm} \sigma_t^{-o}:= f_t(\tau \gamma),\nn
\end{align}

where $f_{t}(\pi_t):= \sum_{k=1}^K f_{tk}(\pi_t)$ is the aggregated inverse marginal utilities adjusted by consumption limits
\bea
f_{tk}(\pi_t):=\max\{0,\min\{L_{tk}^{-1}(\pi_t),\bar{d}_k\}\},\label{eq:ftk}
\eea
and
\begin{equation}\label{eq:edagger}
    \underline{e}^\dagger_t:=\min\{\underline{e},\rho s_t\},~~ \overline{e}^\dagger_t:=\min\{\overline{e},(B-s_t)/\tau\}.
\end{equation}
Given the realized renewable $g_t$ in interval $t$, the optimal storage schedule is
\bea
e^\dagger_t(g_t)=\left\{\begin{array}{ll}
-\underline{e}^\dagger_t, & g_t \le \sigma^{+}_t\\
g_t-\sigma_t^{+o}, & g_t\in (\sigma^{+}_t, \sigma^{+o}_t)\\
0, & g_t \in [\sigma^{+o}_t, \sigma^{-o}_t]\\
 g_t- \sigma_t^{-o}, & g_t \in (\sigma^{-o}_t, \sigma^{-}_t)\\
\overline{e}^\dagger_t, &  g_t \ge \sigma^{-}_t,\\
\end{array}\right.\label{eq:eastThm}
\eea
and the optimal demand schedule, for every $k$, is
\bea 
d_{tk}^\dagger(g_t) = \left\{\begin{array}{ll}
f_{tk}(\pi^+_t), & g_t < \Delta^{+}_t\\
f_{tk}(f^{-1}_t(g_t+\underline{e}_t^\dagger)), & g_t \in [\Delta_{t}^{+},\sigma_{t}^{+}]\\
f_{tk}(f^{-1}_t(\gamma/\rho)), & g_t\in (\sigma^{+}_t, \sigma^{+o}_t)\\
f_{tk}(f^{-1}_t(g_t)), &  g_t \in [\sigma^{+o}_t, \sigma^{-o}_t]\\
f_{tk}(f^{-1}_t(\tau \gamma)), & g_t\in (\sigma^{-o}_t,\sigma_t^{-})\\
f_{tk}(f^{-1}_t(g_t-\overline{e}_t^\dagger)), & g_t\in [\sigma^{-}_t,\Delta_t^{-}]\\
f_{tk}(\pi^-_t), & g_t > \Delta_{t}^{-}.
\end{array}\right. \label{eq:dbf_ast}
\eea
By definition, the total demand is $d_t^\dagger(g_t) := \sum_{k=1}^K d_{tk}^\dagger(g_t)$. 
\end{theorem}
\begin{proof}
    See Appendix~\ref{App:ProofsA}.
\end{proof}

The linear-complexity and closed-form solution in Theorem \ref{thm:Myopic} brings immense intuitions on prosumer response under NEM policies (see Sec.~\ref{sec:prosumer}), and enables ubiquitous DER scheduling, which is explained in Sec.~\ref{subsec:Decentralization}.

\begin{figure}
    \centering
    \includegraphics[scale=0.25]{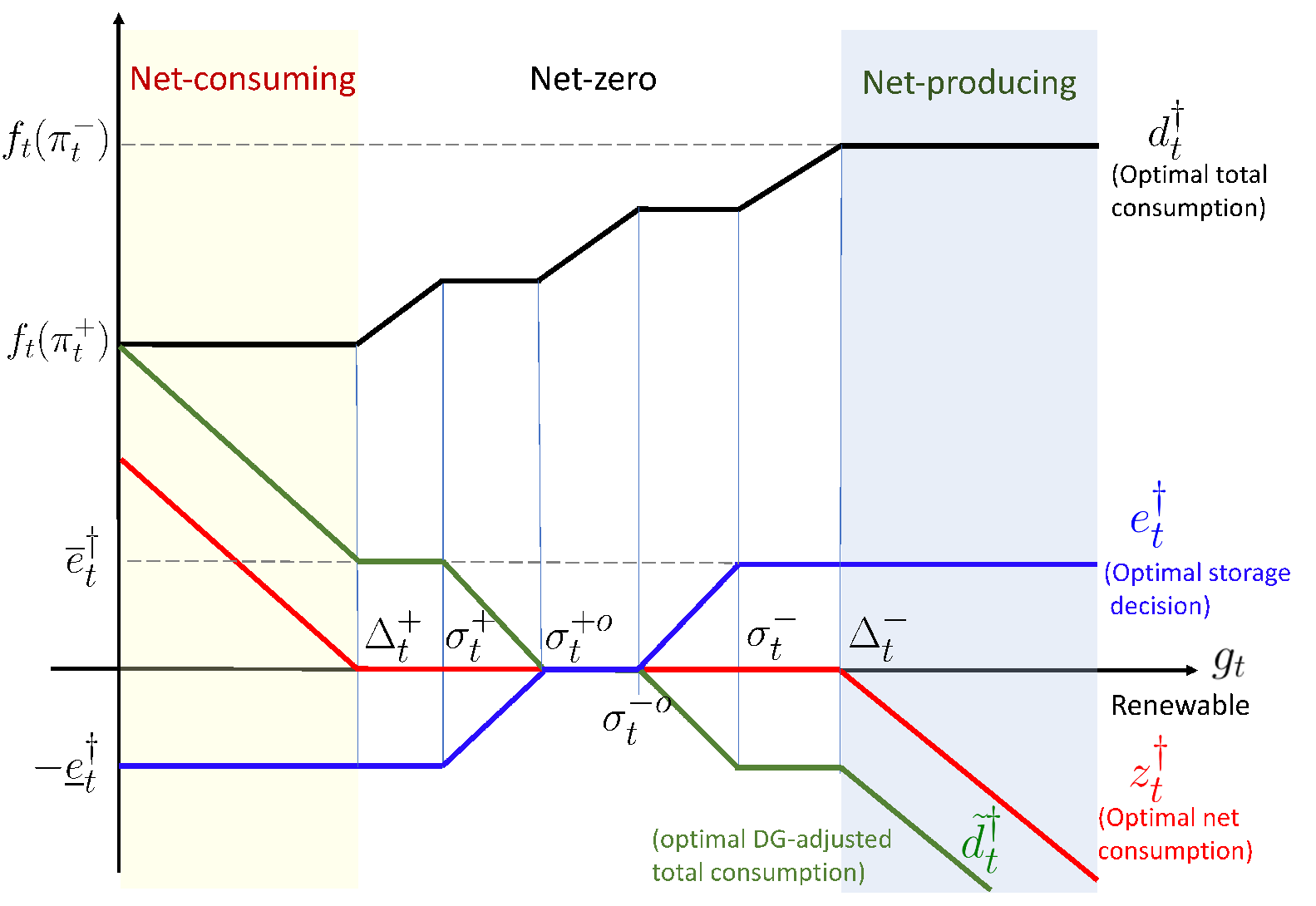}
    \vspace{-0.32cm}
    \caption{Optimal prosumer decisions under the myopic co-optimization: optimal net consumption $z^\dagger_t$ (red),  optimal total consumption $d^\dagger_t$  (black), optimal storage operation $e^\dagger_t$ (blue), and optimal DG-adjusted total consumption $\tilde{d}^\dagger_t:= d_t^\dagger - g_t$ (green), assuming that $\Delta_t^{+}\ge \overline{e}^\dagger_t$ and $\underline{e}^\dagger_t=\overline{e}^\dagger_t$.}
    \label{fig:optdecision}
\end{figure}

Theorem \ref{thm:Myopic} leads to Corollary \ref{corol:NetConsMyopic} below showing that, to the DSO, a prosumer with optimal consumption-storage operation under the myopic co-optimization exhibits three distinct patterns (depicted in Fig.\ref{fig:optdecision}) first observed in \cite{Alahmed&Tong:22TSG} for prosumers with DG and flexible loads, but without BESS: i) net consumption when the renewable is in the net-consuming zone of $(0, \Delta_t^{+})$, ii) net-production when the renewable is in the net-producing zone of $(\Delta_t^{-},\infty)$, and iii) off-the-grid in the net-zero zone of $[\Delta_t^{+},\Delta_t^{-}]$. It is the net-zero zone that is particularly intriguing as it shows the benefit of BTM DER in reducing RPF in the distribution network and the dependency of the prosumer surplus on the NEM X rates, which are further illuminated in Sec.~\ref{sec:num} with simulations. 

\begin{corollary}\label{corol:NetConsMyopic}
    For every $t\in [0,T-1]$, the optimal net consumption $z_t^\dagger$ of the MCO $\Pc_t$ is a monotonically decreasing piecewise-linear function of $g_t$, given by
\begin{equation}\label{eq:OptNetConsumption}
z_t^\dagger(g_t) =\left\{\begin{array}{ll}
\Delta_t^{+} - g_t,   &  g_t < \Delta^{+}_t\\
 0, &  g_t \in [\Delta^+_t, \Delta^{-}_t]\\
\Delta_t^{-} - g_t,   &  g_t > \Delta^{-}_t.\\
\end{array}\right.
\end{equation}
\end{corollary}
\begin{proof}
    See Appendix~\ref{App:ProofsA}.
\end{proof}

As shown in Fig.~\ref{fig:optdecision}, the storage maximally discharges when $g_t \leq  \sigma^{+}_t$, and maximally charges when $g_t \geq  \sigma^{-}_t$, whereas when $g_t\in [\sigma^{+}_t,\sigma^{-}_t]$, the storage output transitions from maximally discharging to maximally charging as $g_t$ increases. When $g_t\in [\sigma^{+o}_t,\sigma^{-o}_t]$, the storage neither charges nor discharges, and the prosumer consumption matches the renewables. Note that $e_t^\dagger$ is determined by only the (inner) thresholds within ($\Delta^{+}_t,\Delta^{-}_t$) as shown in Fig.~\ref{fig:optdecision}.  These thresholds are not affected by NEM X parameters ($\pi_t^+,\pi^-_t$), which means that $e_t^\dagger$ is scheduled independent of ($\pi_t^+,\pi^-_t$). 

Although the proposed linear-complexity solution in Theorem \ref{thm:Myopic} may not be optimal, we demonstrate, in Sec.~\ref{sec:num}, that the solution gap compared to a performance upper bound is relatively small. We also compare the proposed MCO solution with state-of-the-art approaches such as MPC solutions that solve a sequence of rolling-window convex optimization problems with some fixed forecast horizon. The cost of the MPC solution is quite high compared with the MCO scheduling, and it requires renewable forecasts, the quality of which affects its performance. The simulation, in Sec.~\ref{sec:num}, shows that the proposed MCO scheduling algorithm outperformed the MPC solution consistently.

\subsection{Optimality of the MCO Algorithm: a Sufficient Condition}
Theorem \ref{thm:sufficient} below provides a sufficient condition for the proposed MCO algorithm (Theorem \ref{thm:Myopic}) to be optimal. 

\begin{theorem}[Sufficient optimality condition]\label{thm:sufficient}
The MCO algorithm is the optimal solution to the stochastic dynamic program in (\ref{eq:optimization}) if (i) $B > 2T\tau \overline{e}$, and (ii) $s \in(T \underline{e} / \rho, B-T \tau \overline{e})$. 
\end{theorem}
\begin{proof}
    See Appendix~\ref{App:ProofsA}.
\end{proof}

The proof of Theorem \ref{thm:sufficient} leverages Lemmas \ref{prop:MVoS}-\ref{prop:OptS} in the appendix that highlight the structural properties of the optimal prosumer decisions under the sufficient optimality conditions, which are reasonable when battery size $B$ is large or the charging/discharging limits $\overline{e},\underline{e}$ are relatively small. In practice, the two sufficient conditions may not hold in general, but may actually hold if the battery is co-optimized only with a fraction of prosumer flexible loads, making the size to charging limits ratio large. Such practice can be extrapolated to a community-based storage that may serve only a fraction of the community members.
Our numerical results, in Sec.\ref{sec:num}, show that in some operating conditions defined by charging limits and renewable generation levels, the MCO algorithm performs optimally by meeting a theoretical upper bound on total reward, even when the sufficient optimality conditions in Theorem \ref{thm:sufficient} are violated. For example, we found that the algorithm is optimal whenever a storage SoC constraint binds and continue to bind for the rest of the scheduling horizon.

\subsection{Implementation Complexity and Decentralization}\label{subsec:Decentralization}
The closed-form co-optimization solution given in (\ref{eq:DeltaSigma}-\ref{eq:dbf_ast}) has a linear computation cost for the EMS.  Note that functions $f_t,f_{tk}$ in (\ref{eq:ftk}) and thresholds in (\ref{eq:DeltaSigma}) can all be computed offline.  Once the renewable generation $g_t$ is measured, the closed-form expressions of the optimal individual consumption imply that the total computation cost is $\Oc(K)$, \ie linear with respect to the number of energy-consuming devices. The only communication cost comes from communicating $g_t$ to individual devices, which is also of the order $\Oc(K)$ if done sequentially. Note that the implementation of the co-optimization can be easily decentralized.  There is no need to have communications among devices and from devices to the EMS.  The only communication is the broadcast of the level of renewables $g_t$ to all devices. 

It is worth noting that the computation of global thresholds and consumption-storage decisions does not require the knowledge of the underlying probability distribution of $g_t$.   Indeed, the solution given in Theorem \ref{thm:Myopic} applies to arbitrary BTM stochastic generation, which eliminates the need to learn the stochastic model of $g_t$.

%% file: prosumer_v7.tex
In this section, we use the optimal solution under the sufficient optimality condition in Theorem \ref{thm:sufficient} to deliver insights and intuitions on the solution structure. To this end, the optimal solution under the sufficient optimality condition is stated in Corollary \ref{corol:OptDec} next.

\begin{corollary}\label{corol:OptDec}
    For every $t\in [0,T-1]$, under the sufficient optimality conditions in Theorem \ref{thm:sufficient}, the optimal storage operation $e^\ast_t$, consumption $d^\ast_{tk}$, total consumption $d^\ast_t$, and net consumption $z^\ast_t$ are as described in Theorem \ref{thm:Myopic}, but with replacing $\underline{e}_t^\dagger$ and $\overline{e}_t^\dagger$ by $\underline{e}$ and $\overline{e}$, respectively.
\end{corollary}
\begin{proof}
    See Appendix~\ref{App:ProofsA}.
\end{proof}
The implication of Corollary \ref{corol:OptDec} is that the optimal solution to the myopic co-optimization in (\ref{eq:Opt_myopic}) has the same structure as that in the solution of the stochastic dynamic program in (\ref{eq:optimization}) under the sufficient optimality condition. This means that the threshold policy, closed-formedness, and monotonicity properties hold.

\subsection{Optimal Prosumer Decisions: Insights and Intuitions}\label{subsec:Insights}
We consider a special case under the sufficient optimality condition to gain insights into the structure of the optimal prosumer decisions in Corollary \ref{corol:OptDec}, which has the same structure as that in Theorem \ref{thm:Myopic}.

For simplicity, we assume that the prosumer has a single flexible demand ($K=1$) with utility function $U_t(\cdot)$ and marginal utility $L_t(\cdot) = \nabla U_t(\cdot)$. The storage is assumed to be lossless, \ie $\tau=\rho=1$, and the consumption limits constraint in (\ref{subeq:DeviceConstraint}) is ignored.
Let $u_t^\ast:=(d_t^\ast,e_t^\ast)$ be the optimal consumption and storage decisions in interval $t$.

\begin{figure}
    \centering
    \includegraphics[scale=0.25]{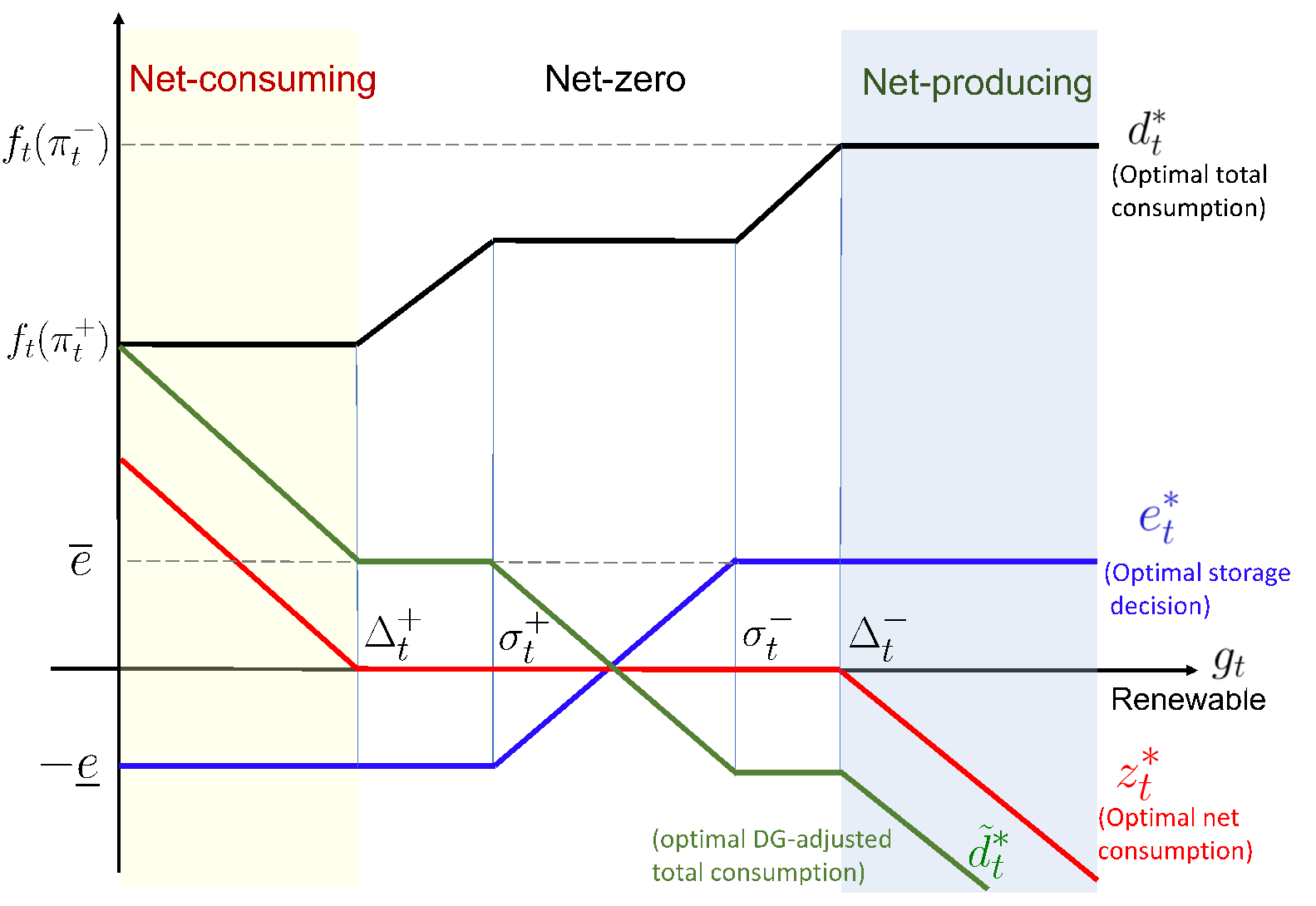}
    \vspace{-0.3cm}
    \caption{Optimal prosumer decisions under special case in Sec.~\ref{subsec:Insights} with decision thresholds:\\
    $\Delta^+_t=\max\{f_t(\pi^+_t)-\underline{e},0\}, \sigma^+_t=\max\{f_t(\gamma)-\underline{e},0\}, \sigma^-_t=f_t(\gamma)+\overline{e}, \\ \Delta^-_t=f_t(\pi^-_t)+\overline{e}$, and assuming $\Delta^+_t \ge \overline{e}$ and $\underline{e}=\overline{e}$.}
        \label{fig:optNetConsdecision}
\end{figure}

The red line in Fig.~\ref{fig:optNetConsdecision} shows the threshold structures in Theorem \ref{thm:Myopic}. The reason behind the three zone structure is that the condition $\pi^-_t \le \gamma \le\pi^+_t$ makes exporting DER to the grid less attractive than using DER to gain more surplus through consumption or storing renewable for future use. The increasing $g_t$ is used with a priority order: 1) offset the consumption to reduce payment, 2) charge the battery, and, 3) export to the grid if there is still excess power left.

The black and blue lines of Fig.~\ref{fig:optNetConsdecision} depicts the optimal consumption and storage decisions, respectively.  The intuitions of these decisions over the three zones are explained next.

\subsubsection[]{Net consumption zone ($z_t^\ast\ge 0$)}  At $g_t=0$, the storage will not charge by Lemma~2.
By Lemma~1, discharging storage to compensate consumption has a marginal value of $\pi^+_t$ higher than keeping it in the storage.  Hence,  it is better to maximally discharge the storage, but not exceed $d_t^\ast$. The optimal consumption is given by
\[
\max_{d_t} \big(U_t(d_t) - \pi^+_t(d_t+e^\ast_t)\big)~\Rightarrow~\left\{\begin{array}{l} d_t^\ast = f_t(\pi^+_t) \\ e^\ast_t=- \min\{ f_t(\pi^+_t), \underline{e}\}.\end{array}\right.
\]


As shown in Fig.~\ref{fig:optNetConsdecision}, as  $g_t$ increases, the renewable $g_t$ and storage are used to reduce payment with the marginal benefits of $\pi^+_t$ until the net consumption $z_t^\ast$ is brought down to zero at $g_t=\Delta^+_t $, which makes $\Delta^+_t=\max\{f_t(\pi^+_t)-\underline{e},0\}$.

\subsubsection[]{Net-production zone ($z_t^\ast\le 0$)}  In this zone, the prosumer is a net-producer facing the compensation rate of $\pi^-_t$.  The optimal consumption is given by
\[
\max_{d_t} \big(U_t(d_t) - \pi^-_t(d_t+e^\ast_t-g_t)\big)~\Rightarrow~\left\{\begin{array}{l} d_t^\ast = f_t(\pi^-_t) \\ e^\ast_t=\min\{ f_t(\pi^-_t), \overline{e}\}.\end{array}\right.
\]
As shown in Fig.~\ref{fig:optNetConsdecision}, as $g_t$ decreases from a very large value, the net consumption $z_t^\ast$ increases to zero at $g_t=\Delta^-_t$, which makes $\Delta^-_t= f_t(\pi^-_t) + \overline{e}$.

\subsubsection[]{Net-zero zone ($z_t^\ast\ = 0$)}   In this zone, there is no payment to/from the utility company. The prosumer schedules its consumption $d_t^\ast$ to maximize its utility subject to that the consumption is offset by a combination of available renewable and storage operation.

Consider the case when $g_t = \Delta^+_t+\epsilon$ for some small $\epsilon>0$.   This $\epsilon$  amount of renewable energy can be used to increase consumption for higher utility (surplus).  While increasing consumption gains in utility, it decreases the marginal utility. So long that $\epsilon$ is small enough that $L_t(f_t(\pi^+_t)+\epsilon) \ge \gamma$,  it is optimal to use the entire $\epsilon$ renewable to increase consumption to $d^\ast_t = f_t(\pi^+_t)+\epsilon$.

When $g_t=\sigma^+_t:= \max\{f_t(\gamma)-\underline{e},0\}$, the marginal value of increasing consumption is equal to charging the storage.  Therefore, for $g_t > \sigma^+_t$, the optimal storage operation $e^\ast_t$  is first reducing the amount of storage discharge followed by increasing level of storage charging until at $g_t=\sigma^-_t  :=f_t(\gamma)+\overline{e}$ when the maximum charging level $\overline{e}$ is reached.  When $g_t > \sigma^-_t$, the prosumer gains further by increasing
its consumption until the marginal utility of consumption matches $\pi^-_t$, at which point $d^\ast_t=f_t(\pi^-_t)$ and $g_t=\Delta^-_t=f_t(\pi^-_t)+\overline{e}$.
When $g_t\ge \Delta^-_t$, the prosumer becomes a net-producer.

Lastly, in the special case of {\em passive prosumers}, under which consumptions are not co-optimized with storage and renewables, we found, in \cite{AlahmedTong:22StorageTPEC}, that the prosumer's optimal policy is to schedule the storage so that it minimizes the absolute value of net-consumption $|z_t|$. 

%% file: num_v7.tex
In this section, numerical analysis is implemented on a standard residential customer under NEM X to show the performance of the myopic co-optimization and the benefits of the co-optimization policy to prosumers and DSOs, over a one-day operation horizon $T=24$ hours starting at midnight. Additional results on the benefits of the co-optimization are relegated to Appendix~\ref{App:ProofsC}. 

To model prosumer's consumption preferences, we adopted a widely-used quadratic concave utility function:
\begin{equation*}
U_{tk}\left(d_{tk}\right)=\left\{\begin{array}{ll}
\alpha_{tk} d_{tk}-\frac{1}{2}\beta_{tk} d_{tk}^2, & 0 \leq d_{tk} \leq \frac{\alpha_{tk}}{\beta_{tk}} \\
\frac{\alpha_{tk}^2}{2 \beta_{tk}}, & d_{tk}>\frac{\alpha_{tk}}{\beta_{tk}}
\end{array}, \forall k,\right.
\end{equation*}
for $t=0,\ldots, T-1$, where $\alpha_{tk}$ and $\beta_{tk}$ are some utility parameters that are learned and calibrated from inputs, including historical prices\footnote{We used PG\&E historical prices, which can be found at \href{https://www.pge.com/tariffs/electric.shtml}{PGE Tariffs}.} and consumption\footnote{The residential load profile data is taken from NREL open dataset for a nominal household in Los Angeles. We used \href{https://shorturl.at/uyL36}{June-August data}.} and by predicating an elasticity for every device type \cite{AlahmedTong:22EIRACM}. Two load types with two different utility functions of the form above were considered: 1) HVAC, 2) other household loads.\footnote{The elasticities of HVAC and other home appliances are taken from \cite{ASADINEJAD_Elasticity:18EPSR}.} We modeled the stochastic household's rooftop solar using the California Solar Initiative's summer months PV profile data with a 5.1 (kW) capacity collected over 2011 to 2016.\footnote{The data can be found at \href{https://www.californiadgstats.ca.gov/downloads}{California DG Statistics.}} The summer period is a more challenging one given that both the consumption, generation, and net consumption profiles are relatively higher.

The prosumer has a battery with specifications similar to Tesla Powerwall,\footnote{The specifications of Tesla Powerwall 2 can be found at \href{https://www.tesla.com/sites/default/files/pdfs/powerwall/Powerwall\%202_AC_Datasheet_en_AU.pdf}{Tesla Powerwall}.} i.e, $B=13.5$ (kWh), and $\tau= \rho= 0.95$. Other battery parameters, including charge/discharge limits, are introduced locally in Sec.~\ref{subsec:numMyopicPerf}--\ref{subsec:numBenefits}.

To model the NEM tariff, we used the retail price $\pi^+$ of the 2020 Pacific Gas and Electric (PG\&E) utility's summer time-of-use (ToU) rates with a 16--21 peak time,\footnote{The PG\&E E-TOU rate (option B) can be found at \href{https://www.pge.com/tariffs/electric.shtml}{PG\&E E-TOU-B}. } which has a peak and offpeak rates of $\pi^+_{\mbox{\tiny ON}}=0.40$ (\$/kWh), $\pi^+_{\mbox{\tiny OFF}}=0.30$ (\$/kWh), respectively. The export rate $\pi^-$ is introduced locally in Sec.~\ref{subsec:numMyopicPerf}--\ref{subsec:numBenefits}. Needless to say, under the regulated retail electricity market, both $\pi^+$ and $\pi^-$ are fixed and known to prosumers {\em apriori}.
The salvage value rate $\gamma$ was set so that (\ref{eq:gamma}) is satisfied.

\vspace{-0.4cm}
\subsection{Performance of Myopic Co-optimization Algorithm}\label{subsec:numMyopicPerf}

We present simulation results that evaluate the performance of the myopic co-optimization (MCO) algorithm developed in Sec.\ref{sec:myopic}.

\subsubsection{Simulation settings}
Starting the daily schedule at midnight, we set the initial storage SoC to empty with $s=0$ (kWh). Setting $s=0$ (kWh) puts the myopic co-optimization algorithm in the disadvantaged position of violating the sufficient optimality condition from the first interval. This setting might not be the case in practice, so we only consider it when we evaluate the performance of the MCO. We consider different, more practical, initial SoC when we study prosumer benefits in Sec.\ref{subsec:numBenefits}. The battery max SoC limit was set to $B=13.5$ (kWh). The export rate was assumed to be $\bm{\pi^-}=0.4 \bm{\pi^+_{\mbox{\tiny OFF}}}$. Using the renewable PV historical data, we compute the hourly mean and variance representing an average summer day. The renewables profile random vector is then generated from a normal distribution with the calculated mean and variance. For MCO, realizations of renewables were generated one sample at a time, and the storage control and consumption decisions were computed and implemented sequentially. Note that the renewables underlying probability distribution is not needed here. For MPC, renewables were also generated one sample at a time and a forecast of renewables in the future $M=4$ intervals was made. An $M$-interval one-shot scheduling optimization was implemented based on a convex relaxation of $\Pc$ in (7), giving a slight advantage of MPC over MCO.  The decision of the immediate scheduling interval was implemented sequentially.

\subsubsection{Performance measure, gap analysis, and benchmark}
The performance measure is the cumulative prosumer reward defined in (\ref{eq:stageReward}). Because the optimal scheduling policy is unknown, we use a theoretical upper bound as the benchmark and define the {\em percentage gap} as a proxy for performance.  

The theoretical upper bound is obtained using a non-causal policy $\mu^\sharp$ that (i) assumes the realization of the renewable for the entire scheduling period, (ii) solves a convex optimization of the one-shot deterministic optimization defined in (\ref{eq:optimization}) relaxed by ignoring simultaneous charging-discharging constraint, and (iii) compute $R_{\mu^\sharp}$ as an upper bound of the cumulative reward. 

We evaluated the performance measured in the percentage gap of MCO with an MPC-based algorithm. Let $R_{\mu^{\mbox{\tiny MCO }}}$ and $R_{\mu^{\mbox{\tiny MPC }}}$ be the cumulative rewards of MCO and MPC over $T$, respectively.  The gap to performance upper bound for each Monte Carlo run is defined by
\begin{equation*}
    G_{\mu^{\mbox{\tiny MCO }}}:=\frac{R_{\mu^{\mbox{\tiny MCO }}} -R_{\mu^{\sharp}}}{R_{\mu^{\sharp}}} \times 100, G_{\mu^{\mbox{\tiny MPC }}}:=\frac{R_{\mu^{\mbox{\tiny MPC }}} -R_{\mu^{\sharp}}}{R_{\mu^{\sharp}}} \times 100.
\end{equation*}
Note that if the gap is zero, the algorithm is optimal.  When it is not, the performance may or may not be optimal.

\subsubsection{Results and discussions}
\begin{figure}
    \centering
    \includegraphics[scale=0.27]{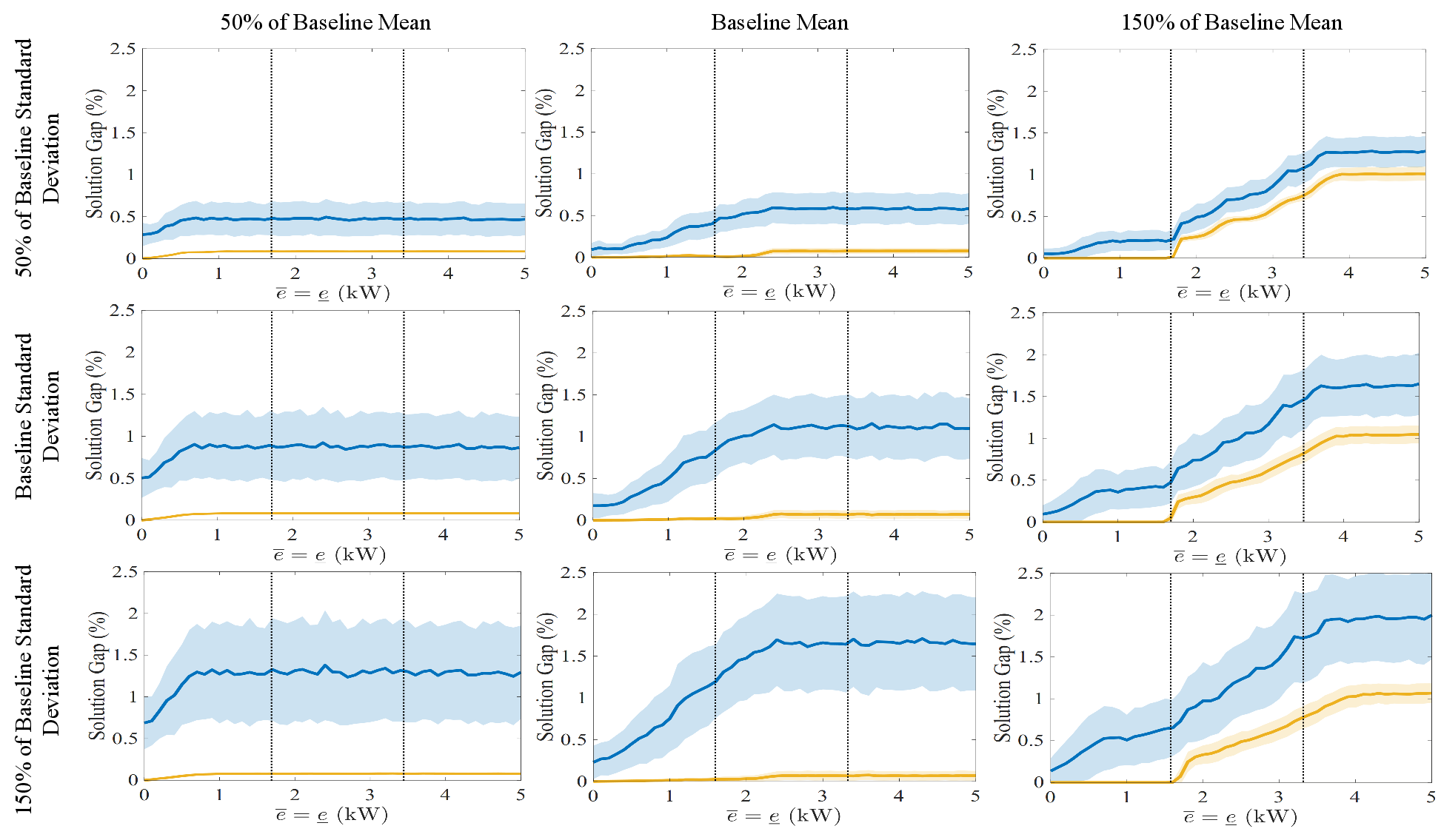}
    \vspace{-0.7cm}
    \caption{Solution gap of MCO and MPC algorithms under different renewable levels. The two vertical lines represent the $C-8$ and $C-4$ storage charging/discharging levels.}
    \label{fig:SolGap}
\end{figure}

Using 500 Monte Carlo runs on the renewable DG, Fig.\ref{fig:SolGap} shows the average solution gap of the MCO algorithm (yellow) and the MPC algorithm (blue) compared to the upper bound under varying storage charging/discharging rates. The columns, from left to right, show the solution gap under 50\% of baseline mean of renewables (i.e., the mean of each interval is 50\% the mean of baseline), baseline mean of renewables (i.e., from historical data), and 150\% of baseline. The rows, from top to bottom, show the solution gap under 50\% of baseline standard deviation of renewables (i.e., the standard deviation of each interval is 50\% the standard deviation of baseline), baseline standard deviation of renewables, and 150\% of baseline.

In all renewable level cases, MCO outperformed MPC. For the practical battery setting---the 8-hour ($C-8$) and 4-hour ($C-4$) charging cases, and for all renewable level cases, MCO achieved a solution gap that did not exceed 0.75\% well within the practically acceptable performance levels. Also in all renewable level cases, the solution gap saturated after a certain charging/discharging power. In both the baseline and 50\% of baseline mean renewable levels under baseline and 150\% of baseline standard deviation, the solution gap under MPC is on average more than 15 times greater than it is under MCO. In the 150\% of baseline mean renewable case and for all standard deviation levels, MCO's solution gap was zero when the charge/discharge power was less than $\sim$1.7 (kW). The optimality of MCO below $\sim$1.7 (kW) is because the higher renewable level managed to adequately increase the SoC in the midday, ensuring that the minimum SoC limit is not reached again (given $\overline{e}=\underline{e}$) when the renewable output fades out. As the charging/discharging power increased, the prosumer was able to reach maximum SoC of $B$ during midday, or minimum SoC of $0$ later in the day, or both. The zero gap when $\overline{e}=\underline{e}<1.7$ (kW) shows, one of the scenarios whereby MCO is optimal even though the sufficient optimality condition was violated. 

Over all cases, the solution gap of both algorithms increased when the mean and standard deviation of renewables were increased. For instance, when the standard deviation of renewables increased from 50\% to 150\% of baseline under the baseline mean case, the MCO's solution gap increased from 0.07\% to 0.075\% and the MPC's solution gap increased from 0.59\% to 1.65\% when the charging/discharging power was at ($C-4$). This is because when the mean was higher, the probability of violating the SoC limits increased. The same reasoning follows when the standard deviation was increased.

\subsection{Computational Efficiency Comparison}\label{subsec:numMyopicCompu}
In Table \ref{tab:MCOcomputations}, we compare the computational efficiency of MCO and MPC algorithms over the one-day operation horizon. The second and third columns in Table \ref{tab:MCOcomputations}, respectively, show the computation time in seconds and the worst-case run time of the MCO and MPC algorithms when $K=1$ explicitly by taking use of the discrete structure of the optimization problem.

As shown in Table \ref{tab:MCOcomputations}, for our relatively small problem of $24$ intervals, MPC's computations with 4 hours lookahead were 170 times longer than MCO's computation time. Increasing the lookahead window to $M=12$ increased the computation time ratio with MCO to 192. This ratio would be even higher if a more complex renewable forecasting approach was adopted \cite{Hoangetal:TPS24}. The computational complexity of the MCO grows linearly with $T$ \ie $\Oc(T)$, whereas, under MPC, it grows at least to the fourth power with the size of the lookahead window, \ie $\Oc(T (M^4))$. The computational complexity is higher when the utility's quadratic functional form is replaced with the more general class of concave functions.

\begin{table}
\centering
\caption{Computational efficiency.$^\ddagger$}
\vspace{-0.27cm}
\label{tab:MCOcomputations}
\resizebox{0.6\columnwidth}{!}{%
\begin{tabular}{@{}ccc@{}}
\toprule \midrule
Algorithm   & Computation Time (s) \\ \midrule
MCO         &      0.03      \\
MPC ($M=4$) &     5.1       \\
MPC ($M=8$) &      5.43      \\
MPC ($M=12$) &      5.75       \\
\midrule \bottomrule
\end{tabular}%
}
\begin{adjustwidth}{0pt}{0pt} \footnotesize{ \begin{flushleft} \vspace{0.2cm} {\tiny $^\ddagger$Using a machine with a 64-bit Intel(R) Xeon(R) Gold 6230R CPU @ 2.10 GHz processor and 256GB installed RAM. For MPC, YALMIP \cite{Lofberg:CACSD04} with Gurobi was used.}
 \end{flushleft}}
\end{adjustwidth}
\end{table}

\subsection{Prosumer Benefits}\label{subsec:numBenefits}
We present here the prosumer benefits achieved by the co-optimization policy and other scheduling policies.

\subsubsection{Simulation Settings} 
Five customer types are considered. (1) {\em Consumers}: customers without BTM DER. (2) {\em Solar exporters}: prosumers who discharge the storage during ToU peak hours to match consumption, in order to sell back the renewable generation. This is similar to Tesla Powerwall's {\em solar energy exports} mode. (3) {\em Self-powered} customers: prosumers who, for every $t$, schedule their storage to negate the DG-adjusted consumption $\tilde{d}_t$ \cite{AlahmedTong:22StorageTPEC}. This is also Tesla Powerwall's {\em self-powered} mode. (4) {\em Active SDG} customers: prosumers who co-optimize storage and consumption based on available BTM DG according to MCO. (5) {\em Packaged SDG} customers: prosumers who emulate the case when DG and storage are combined in a single unit whose control prioritizes using renewables to charge the storage before meeting the consumption. Specifically, if 
\begin{enumerate}
    \item $g_t>0$: the storage charges $e^\ast_t=\min(g_t,\overline{e})$ and $d^\ast_{tk}$ for every $k$ is as in \cite{Alahmed&Tong:22TSG}, but with $\tilde{g}_t:=g_t-e_t$ instead of $g_t$.
    \item $g_t=0$: the optimal policy co-optimizing the storage
and consumption is as in Theorem \ref{thm:Myopic}.
\end{enumerate}

Throughout the analysis, the initial storage SoC $s$ was set at $s= 12.15$ (kWh), and the battery SoC limits were set at $B=12.83$ (kWh), and $\underline{B}=0.68$ (kWh).\footnote{Here we consider a more general setting where the SoC constraints are given by $\underline{B} \le s_t \le B$.}

\subsubsection{Prosumer surplus}

We compared the cumulative rewards achieved by the five customer types. Using {\em consumers} as the reference, Table \ref{tab:RewardToGo} shows the percentage gain in total surplus under different storage charging/discharging powers and two export rates\footnote{The social marginal cost (SMC) rate $\pi^{\mbox{\tiny SMC}}$ in Table \ref{tab:RewardToGo} is as discussed in \cite{NEMevolution:23NAS}, which
is slightly above the wholesale price. The day-ahead LMP data is taken from CAISO SP15 for the period June-August, 2019 \href{http://oasis.caiso.com/mrioasis/logon.do}{(OASIS-CAISO)}.} over that achieved by a consumer. 

The highest surplus was achieved by active SDG prosumers, as they were the most effective in avoiding buying at $\bm{\pi^+}$ and being compensated at $\bm{\pi^-}$. The active SDG prosumer, under $\overline{e}=\underline{e}=1.5$ (kW) and $\bm{\pi^-} = \bm{\pi^{\mbox{\tiny SMC}}}$, achieved a cumulative surplus gain of 102\%, whereas solar exporter, self-powered prosumer, and packaged SDG prosumer have only gained 83.5\%, 95.8\%, and 87.7\%, respectively. Table \ref{tab:RewardToGo} shows that reducing $\bm{\pi^-}$ decreased all surpluses, but increased the surplus differences between active SDG and other prosumers because the consumption of active SDG dynamically increases to avoid/reduce net exports at the decreased sell rate. The cumulative surplus gain differences between active SDG and self-powered, and solar export prosumers were 3.7\% and 7.3\%, respectively, under $\bm{\pi^-}=0.6\bm{\pi^+_{\mbox{\tiny OFF}}}$ and $\overline{e}=\underline{e}=1.5$ (kW), which increased to 6.2\% and 18.5\% when the sell rate was reduced to $\bm{\pi^-} = \bm{\pi^{\mbox{\tiny SMC}}} \preceq 0.6\bm{\pi^+_{\mbox{\tiny OFF}}}$. 

When the storage output was increased (from 0.5 (kW) to 1.5 (kW)), the optimality of active SDG became more apparent, and the influence of $\bm{\pi^-}$ diminished as more generation was kept behind the meter. The active SDG prosumer's cumulative surplus gain, under $\overline{e}=\underline{e}=0.5$ (kW), increased by 9.6\% when the export rate increased to $0.6\bm{\pi^+}$, but when $\overline{e}=\underline{e}=1.5$ (kW), the surplus gain increase was only 1.9\%.

\begin{table}
\centering
\vspace{-0.3cm}
\caption{cumulative surplus gain (\%) over consumer's surplus}
\vspace{-0.3cm}
\label{tab:RewardToGo}
\resizebox{0.49\textwidth}{!}{%
\begin{tabular}{@{}ccccccc@{}}
\toprule \toprule
\multirow{2}{*}{\begin{tabular}[c]{@{}c@{}}$\bm{\pi^-}$\\ (\$/kWh)\end{tabular}} & \multirow{2}{*}{\begin{tabular}[c]{@{}c@{}}$\overline{e}=\underline{e}$\\ (kW)\end{tabular}} & \multicolumn{4}{c}{Customer Type} \\
 &  & Consumer & Solar exporter & Self-powered & Packaged SDG & Active SDG \\ \midrule
\multirow{3}{*}{$\bm{\pi^{\mbox{\tiny SMC}}}$} & 0.5 & 0\%  & 68.2\% & 72.9\% & 77.9\%& \textbf{83.3}\% \\
 & 1 & 0\%  & 76.7\% & 85.7\% & 84.1\% & \textbf{94.3}\% \\
 & 1.5 & 0\%  & 83.5\% & 95.8\% & 87.7\% & \textbf{102.0}\% \\ \midrule
\multirow{3}{*}{$0.6\bm{\pi^+_{\mbox{\tiny OFF}}}$} & 0.5 & 0\%  & 87.9\% & 89.4\% & 87.4\% & \textbf{92.9}\% \\
 & 1 & 0\%  & 92.9\% & 95.9\% & 89.0\% & \textbf{99.2}\% \\
 & 1.5 & 0\%  & 96.6\% & 100.2\% & 89.5\% & \textbf{103.9}\% \\ \bottomrule \bottomrule
\end{tabular}%
}
\end{table}

\begin{figure}
\vspace{-0.29cm}
    \centering
    \includegraphics[scale=0.35]{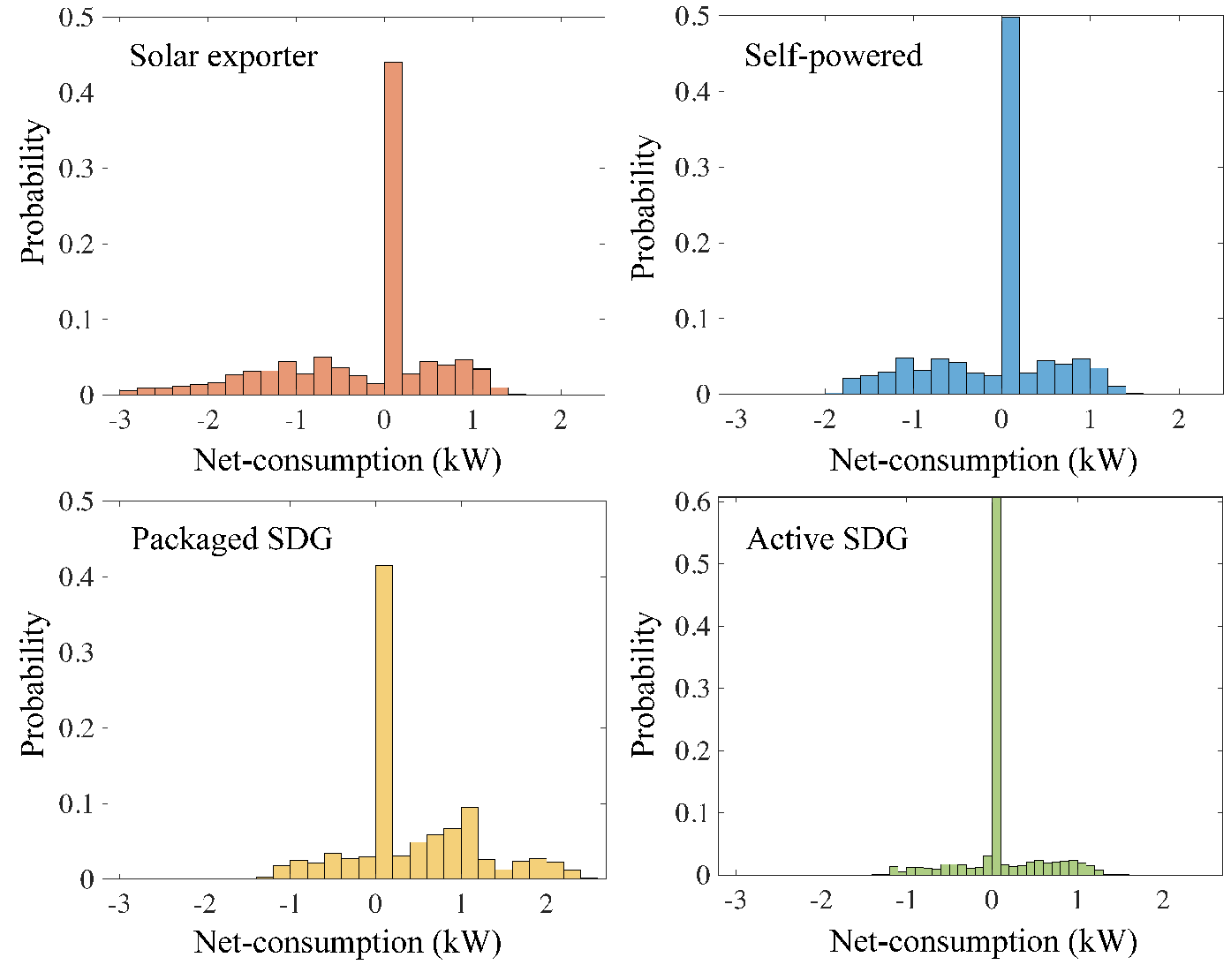}
    \vspace{-0.35cm}
    \caption{Net consumption distribution ($\overline{e}=\underline{e}=1$ (kW)).}
    \label{fig:OptNetCons}
    \vspace{-0.4cm}
\end{figure}

We use the net consumption distribution of different customer types in Fig.\ref{fig:OptNetCons} to expand on the intuition behind Table \ref{tab:RewardToGo}. The distribution was acquired by implementing 50,000 Monte Carlo runs on renewable DG. Fig.\ref{fig:OptNetCons} shows that, given the high amount of exported power of solar exporters (top left), their cumulative surplus was heavily influenced by $\bm{\pi^-}$ (see Table \ref{tab:RewardToGo}). Self-powered prosumers (top right) had less energy imports probability than packaged SDG prosumers (bottom left), but higher exports probability, because they avoided buying from the grid by using their renewables and storage to satisfy the demand. Their energy exports are higher than packaged and active SDG prosumers because only the storage reacts to absorb excess generation. The packaged SDG prosumers' policy of prioritizing charging the storage from DG and increasing consumption to absorb $\tilde{g}_t$ reduced net exports, but not net imports. Lastly, the active SDG prosumer's co-optimization of storage and consumption (bottom right) noticeably centered the net consumption around the origin (zero), showing the co-optimization efficacy in putting the prosumer off the grid. Higher net-zero zone probability enabled active SDG prosumers to avoid buying at $\bm{\pi^+}$ or selling at $\bm{\pi^-}$, resulting in the high surplus gain (Table \ref{tab:RewardToGo}).

%% file: appendix_Nomenclature.tex
\addcontentsline{toc}{section}{Nomenclature}
\subsection*{Abbreviations}
\begin{IEEEdescription}[\IEEEusemathlabelsep\IEEEsetlabelwidth{$V_1,V_2,V_3$}]
\item[BTM] Behind the meter.
\item[DER] Distributed energy resources.
\item[DG] Distributed generation.
\item[DSO] Distribution system operator.
\item[EMS] Energy management system
\item[MCO] Myopic co-optimization.
\item[MDP] Markov decision process.
\item[MPC] Model predictive control.
\item[NEM] Net energy metering.
\item[RPF] Reverse power flows.
\item[SoC] State of charge.
\end{IEEEdescription}
\subsection*{Parameters and Indices}
\begin{IEEEdescription}[\IEEEusemathlabelsep\IEEEsetlabelwidth{$V_1,V_2,V_3$}]
\item[$B \in \mbbR_+$] Storage SoC upper limit.
\item[$\bm{\overline{d}} \in \mbbR_+^K$] Consumption bundle’s upper limit.
\item[$\overline{e},\underline{e}\in \mbbR_+$] Storage charging and discharging limits.
\item[$\gamma \in \mbbR_+$] Salvage (marginal) value of storage energy.
\item[$k,K$] Device index and total number of devices.
\item[$\pi^+_t, \pi^-_t \in \mbbR_+$] ~~NEM buy and export rates.
\item[$\rho, \tau \in \mbbR_+$] Storage discharging and charging efficiencies.
\item[$t,T$] Time index and total number of intervals.
\end{IEEEdescription}
\subsection*{Variables and Functions}
\begin{IEEEdescription}[\IEEEusemathlabelsep\IEEEsetlabelwidth{$V_1,V_2,V_3$}]
\item[$\bm{d},\bm{d}^\ast$] Consumption and optimal consumption bundles.
\item[$d_t,d_t^\ast$] Total consumption and optimal total consumption. 
\item[$d_{tk},d_{tk}^\ast$] $k$th device’s consumption and optimal consumption.
\item[$\tilde{d}_{t},\tilde{d}_{t}^\ast$] DG-adjusted and optimal DG-adjusted total consumption.
\item[$e_t,e_t^\ast$] Storage and optimal storage control. 
\item[$f_{tk}$] Inverse marginal utility of device $k$. 
\item[$g_t \in \mbbR_+$] BTM renewable generation.
\item[$L_{tk}(\cdot)$] Marginal utility function of device $k$. 
\item[$\mu_t$] Decision rule of an MDP policy.
\item[$P^{\mbox{\tiny NEM}}_{\pi_t}(\cdot)$] Payment function under NEM tariff $\pi$.
\item[$r_t(\cdot)$] Stage reward function.
\item[$s_t$] Storage SoC in interval $t$.
\item[$S^{\mbox{\tiny NEM}}_{\pi_t}(\cdot)$] Surplus function under NEM tariff $\pi$.
\item[$u_t$] Control action of the MDP.
\item[$U_{tk}(\cdot)$] Utility function of device $k$.
\item[$V(\cdot)$] Optimal reward-to-go function.
\item[$x_t$] State of the MDP.
\item[$z_t,z_t^\ast$] Net-consumption and optimal net-consumption.
\end{IEEEdescription}

%% file: AppendixA_v7.tex
\subsection[]{Preliminary: Prosumer Without Storage ($\overline{e}=\underline{e}=0$)}  Without storage, the optimal consumption of a prosumer is a static optimization.  The optimal consumption is shown to have a two-threshold three-zone structure.

The following Theorem is shown in \cite{Alahmed&Tong:22TSG}.

\begin{theorem}[Prosumer consumption decision under NEM X]\label{thm:OptD}
When $\overline{e}=\underline{e}=0$, the optimal prosumer consumption policy under $\max\{(\pi_t^-)\} \leq \min\{(\pi_t^+)\}$, for every $t$ is given by two thresholds
\beq
\Delta_t^+:=f_t(\pi^+_t),~~\Delta^-_t:= f_t(\pi^-_t)
\eeq
that partition the range of DER production into three zones:
\ben
\item \underline{Net consumption zone:  $g_t<\Delta^+_t$}. The prosumer is a net-consumer with consumption
\beq \label{eq:d_i^+}
d^\ast_{tk} = f_{tk}(\pi^+_t),~~\forall k.
\eeq
\item  \underline{Net producing zone:  $g_t>\Delta_t^-$}. The prosumer is a net-producer with consumption
\beq \label{eq:d_i^-}
d^\ast_{tk} = f_{tk}(\pi^-_t),~~\forall k.
\eeq
\item  \underline{Net-zero energy zone:  $\Delta^+_t \le g_t \le \Delta^-_t$}. The prosumer is a net-zero consumer with consumption:
\beq \label{eq:d_i^o}
d_{tk}^\ast = f_{tk}(f^{-1}_t (g_t)),~~\forall k.
\eeq
\een
\end{theorem}


\subsection[]{Preliminary: Lemmas \ref{prop:MVoS}--\ref{prop:OptS}}

\input{appendix_lemmas_v7}


\subsection{Proof of Theorem \ref{thm:sufficient}}
We leverage Lemmas \ref{prop:MVoS}--\ref{prop:OptS} to prove the optimality of the MCO under the sufficient conditions: (i) $B > 2T\tau \overline{e}$, and (ii) $s \in(T \underline{e} / \rho, B-T \tau \overline{e})$, by deriving first the optimal net consumption $z^\ast_t$ (part I below) and then the optimal consumption $d^\ast_t$ and storage operation $e^\ast_t$ (part II below), followed by showing that these values match the MCO's $z^\dagger_t$, $d^\dagger_t$, and $e^\dagger_t$, respectively. Note that under the sufficient optimality conditions, $\underline{e}^\dagger_t = \underline{e}$ and $\overline{e}^\dagger_t = \overline{e}$, for all $t\in [0,T-1]$.

-- \textit{\underline{Part I (optimal net consumption $z^\ast_t$ and monotonicity}} \textit{\underline{of $d^\ast_t$ and $e^\ast_t$)}:}

By Lemma \ref{prop:OptS}, we have $e_t^\ast \le g_t$.
Let the storage-augmented renewable be $y^\ast_t:=g_t-e_t^\ast \ge 0$.

From Theorem 1 in \cite{Alahmed&Tong:22TSG}, we have
\begin{equation} \label{eq:dastPf}
d_t^\ast= \begin{cases}f_t\left(\pi_t^{+}\right), & y_t^\ast<f_t\left(\pi_t^{+}\right) \\ y_t^\ast, & y_t^\ast \in \left[f_t\left(\pi_t^{+}\right), f_t\left(\pi_t^{-}\right)\right] \\ f_t\left(\pi_t^{-}\right), & y_t^\ast>f_t\left(\pi_t^{-}\right) .\end{cases}
\end{equation}
where $f_t:=\sum_{k=1}^K f_{tk}$.

With the optimal net-consumption $z_t^\ast=d^\ast_t+e^\ast_t-g_t$,

\begin{equation*}
z_t^\ast=\left\{\begin{array}{l}
f_t\left(\pi_t^{+}\right)+e_t^\ast-g_t>0, \quad g_t<f_t\left(\pi_t^{+}\right)+e_t^\ast \\
0, \quad f_t\left(\pi_t^{+}\right)+e_t^\ast \leq g_t \leq f_t\left(\pi_t^{-}\right)+e_t^\ast \\
f_t\left(\pi_t^{-}\right)+e_t^\ast-g_t<0, \quad g_t>f_t\left(\pi_t^{-}\right)+e_t^\ast
\end{array}\right.
\end{equation*}

\underline{Case I ($g_t<\Delta_t^+$):}  With $e_t^\ast \ge -\underline{e}$, we have
\[
g_t < \Delta_t^+ \le f_t(\pi^+_t)+e_t^\ast~~\Rightarrow~~y_t^\ast:=g_t-e^\ast_t < f_t(\pi^+_t).
\]
From (\ref{eq:dastPf}), $d^\ast_t=f_t(\pi^+_t)$ when $g_t < \Delta_t^+=d^\ast_t-\underline{e}$.
From Lemma~\ref{prop:OptS}, $e_t^\ast=\max\{g_t-d_t^\ast,-\underline{e}\}=\underline{e}$. Therefore, when $g_t < \Delta_t^+$, $z^\ast_t = d^\ast_t+e^\ast_t-g_t = \Delta_t^+-g_t.$

\underline{Case II ($g_t>\Delta_t^-$):}  With $e_t^\ast \le \overline{e}$, we have
\[
g_t > \Delta_t^- \ge f_t(\pi^-_t)+e_t^\ast~~\Rightarrow~~y_t^\ast:=g_t-e^\ast_t > f_t(\pi^-_t).
\]
From (\ref{eq:dastPf}), $d^\ast_t=f_t(\pi^-_t)$ when $g_t > \Delta_t^-=d_t^\ast + \overline{e}$.
From Lemma~\ref{prop:OptS}, $e_t^\ast=\min\{g_t-d_t^\ast,\overline{e}\}=\overline{e}$. Therefore, when $g_t > \Delta_t^+$, $z^\ast_t = d^\ast_t+e^\ast_t-g_t = \Delta_t^--g_t.$

\underline{Case III ($g_t \in [\Delta_t^+,\Delta_t^-]$):}  Note that at the two ends of the interval
$z_t^\ast(g_t)$ matches at $z_t(\Delta_t^+)=z_t^\ast(\Delta_t^-)=0$.  To show $z_t^\ast(g_t)=0$ in Case III, we only need to argue that $z_t^\ast$ must be a monotonically decreasing with $g_t$.

Consider $z_t^\ast=0$ at $g_t$.
For $\epsilon>0$, let $g_\epsilon:=g_t+\epsilon$, denote the optimal control action under $g_\epsilon$ by  $u^\ast_\epsilon=(e^\ast_\epsilon, d^\ast_\epsilon)$.  Let
$z^\ast_\epsilon=d_\epsilon+e_\epsilon-g_\epsilon$ be the optimal net-consumption.

Suppose that
\beq \label{eq:zstarZero}
z^\ast_\epsilon>0~~\Rightarrow~~d^\ast_\epsilon > g_\epsilon - e^\ast_\epsilon.
\eeq
By Lemma~\ref{prop:complementarity}, $e^\ast_\epsilon \le 0~\Rightarrow~d^\ast_\epsilon > g_\epsilon$.  By Lemma~\ref{prop:OptS},
\[
e^\ast_\epsilon = \max\{g_t-d^\ast_t, -\underline{e}\} = -\underline{e}
\]
Let $y^\ast_\epsilon:=g_\epsilon-e^\ast_\epsilon= g_\epsilon+\underline{e}$, we have
\[
y^\ast_\epsilon \stackrel{(a)}{\ge} y^\ast_t:=g_t - e^\ast_t \stackrel{(b)}{=} d^\ast_t \stackrel{(c)}{=} f_t(\pi^+),
\]
where  we use $e_t^\ast \ge -\underline{e}$ and $g_\epsilon>g_t$ in (a), $z_t^\ast=0$ in (b), and  (\ref{eq:dastPf}) in (c).  Again from (\ref{eq:dastPf}), we have $d_\epsilon^\ast \le y^\ast_\epsilon$ whenever $y^\ast_\epsilon > f_t(\pi^+_t)$. Therefore,
\[
z_\epsilon^\ast = d_\epsilon^\ast + e_\epsilon^\ast - g_\epsilon \le y_\epsilon^\ast + e_\epsilon^\ast - g_\epsilon = 0.\]
Contradiction to (\ref{eq:zstarZero})

Hence, combining cases I--III, we have
\begin{equation}\label{eq:OptNetConsumptionNSL}
z_t^\ast(g_t) =\left\{\begin{array}{ll}
\Delta_t^{+} - g_t,   &  g_t < \Delta^{+}_t\\
 0, &  g_t \in [\Delta^+_t, \Delta^{-}_t]\\
\Delta_t^{-} - g_t,   &  g_t > \Delta^{-}_t.\\
\end{array}\right.
\end{equation}

To show that $d_t^\ast$ is monotonically increasing with $g_t$, note that we have already shown that $d_t^\ast=f_t(\pi_t^+)$ when $g_t \le \Delta^+_t$, and $d_t^\ast=f_t(\pi^-_t) \ge f_t(\pi_t^+)$ when $g_t \ge \Delta_t^-$.  We only need to show that, when $g_t \in (\Delta_t^+,\Delta_t^-)$, $d_t^\ast$ is monotonically increasing.

Given state $x_t=(s_t,g_t)$ at time $t$, the optimization in interval $t$ is given by
\beq \label{eq:opt_zero}
\begin{array}{ll}
\underset{(d_{tk}),e_t}{\mbox{maximize}} & \sum_k U_{tk}(d_{tk}) + \gamma(\tau [e_t]^+-\frac{1}{\rho}[e_t]^-)\\
\mbox{subject to} & \\
(\lambda): & \sum_k d_{tk} + [e_t]^+-[e_t]^- = g_t\\
(\underline{\mu}^+,\bar{\mu}^+): & 0 \le [e_t]^+ \le \overline{e}\\
(\underline{\mu}^-,\bar{\mu}^-): & 0 \le [e_t]^- \le \underline{e}\\
(\underline{\eta}_k,\bar{\eta}_k): & 0 \le d_{tk} \le \bar{d}_{tk}\\
\end{array}
\eeq
where the objective function comes from the Bellman's equation and Lemma~\ref{prop:MVoS}, and the equality constraint from $z_t^\ast=0$.

From the KKT conditions, we have
\bea
L_{tk}(d^\ast_{tk})&=& \lambda^\ast-\bar{\eta}^\ast_k +\underline{\eta}_k^\ast \label{eq:Lk}\\
g_t &=& \sum_k d_{tk}^\ast + [e_t]^{+^\ast} - [e_t]^{-^\ast}. \label{eq:gt}
\eea

We consider constraint binding scenarios:
\ben
\item $e^\ast_t=\overline{e}$: In this case, $[e_t]^{+^\ast} = \overline{e}, [e_t]^{-^\ast}=0$.  From (\ref{eq:gt}), the total demand grows {\em linearly} $g_t$.
\item $0 < e^\ast_t < \overline{e}$:  In this case, $(\bar{\mu}^{+*},\underline{\mu}^{+*},\bar{\mu}^{-*},\underline{\mu}^{-*})={\bm 0}$.
By the KKT condition, $\lambda^\ast=\gamma\tau$ is independent of $g_t$.  Therefore,
\[
d_{tk}^\ast=\min\{f_{tk}(\lambda^\ast),\bar{d}_{tk}\},
\]
which is independent of $g_t$.  By (\ref{eq:gt}) $e^\ast_t$ grows {\em linearly} with $g_t$.
\item $ e^\ast_t =0$:  In this case, $\sum_k d_{tk}^\ast=g_t$, making total consumption grow linearly  with $g_t$.
\een

The proof of monotonicity for the  case $-\underline{e} \le e^\ast_t < 0$ follows the same arguments above.

Notice that, for both total the optimal consumption $d_{tk}^\ast$ and storage operation $e_t^\ast$, they either hold constant or growing linearly with $g_t$ at the slope of one, which implies that both are monotonically increasing piecewise linear function.

-- \textit{\underline{Part II (optimal consumption $d^\ast_t$ and storage operation $e^\ast_t$)}:}
We consider decisions in each interval shown in Fig.~\ref{fig:optdecision}, leveraging  Part I above and Lemmas \ref{prop:MVoS}-\ref{prop:OptS}.

\underline{Case I ($ 0 \le g_t \le \Delta_t^+$):} From Part I, we have $z_t^\ast\ge 0$.  By the complementarity condition of Lemma~\ref{prop:complementarity}, we have $e_t^\ast\le 0$.  Let $y_t^\ast:=g_t-e_t^\ast\ge 0$ be the storage-augmented renewable. Because $g_t \le f_t(\pi_t^+)-\underline{e}$,
\[
y_t^\ast \le f_t(\pi^+_t)-(\underline{e} + e_t^\ast) \le f_t(\pi^+_t).
\]
From Theorem 1 in \cite{Alahmed&Tong:22TSG}, when $y_t^\ast \le f_t(\pi^+_t)$, the optimal consumptions of device $k$ is given by
\bea
d_{tk}^\ast &=& f_{tk}(\pi_t^+),~~\mbox{for all $k$.}\nn\\
d_t^\ast &=& \sum_k d_{tk}^\ast=f_t(\pi^+_t).\nn
\eea
From Lemma~\ref{prop:OptS}, with $g_t\le d_t^\ast - \underline{e}$, we have $e_t^\ast=-\underline{e}$.

\underline{Case II ($g_t \ge \Delta_t^-$):}
From Part I, we have $z_t^\ast < 0$.  By the complementarity condition of Lemma~\ref{prop:complementarity}, we have $e_t^\ast \ge 0$ and, by Lemma~\ref{prop:OptS}, $e_t^\ast\le g_t$, and the storage-augmented renewable
$y_t^\ast:=g_t-e_t^\ast\ge 0$. Because $g_t > f_t(\pi_t^-)+\overline{e}$,
\[
y_t^\ast \ge f_t(\pi^-_t)+\overline{e} - e_t^\ast \le f_t(\pi^-_t).
\]
From Theorem 1 in \cite{Alahmed&Tong:22TSG}, when $y_t^\ast\le f_t(\pi^+_t)$, the optimal consumptions of device $k$ is given by
\bea
d_{tk}^\ast &=& f_{tk}(\pi_t^-),~~\mbox{for all $k$.}\nn\\
d_t^\ast &=& \sum_k d_{tk}^\ast=f_t(\pi^-_t).\nn
\eea
From Lemma~\ref{prop:OptS}, with $g_t\ge d_t^\ast + \overline{e}$, we have $e_t^\ast=\overline{e}$.

\underline{Case III ($\Delta_t^+ < g_t < \Delta^-_t$):}  This is the case in the net-zero zone where $z_t^\ast=0$, and the optimization involved is defined by (\ref{eq:opt_zero}).

From Part I, we have shown that both the total consumption $d_t^\ast$ and storage operation $e^\ast_t$ are both piecewise linear monotonically increasing functions of $g_t$.  Furthermore, whenever $e_t^\ast$ strictly increasing with $g_t$, $d_{t}^\ast$ holds constant.

\underline{Case III-a ($\Delta_t^+ < g_t \le \sigma^+_t$):} We check the KKT condition for the case that $e_t^\ast=-\underline{e}$.  The optimization (\ref{eq:opt_zero}) reduces to
\beq \label{eq:opt_CaseIII-a}
\begin{array}{ll}
\underset{(d_{tk})}{\mbox{maximize}} & \sum_k U_{tk}(d_{tk})\\
\mbox{subject to} & \\
(\lambda): & \sum_k d_{tk}  = g_t + \underline{e}\\
(\underline{\eta}_k,\bar{\eta}_k): & 0 \le d_{tk} \le \bar{d}_{tk},\\
\end{array}
\eeq
from which we have $d_{tk}^\ast=f_{tk}(f_t^{-1}(g_t+\underline{e}))$.

\underline{Case III-b ($\sigma_t^+ < g_t < \sigma^{+o}_t$):}
 We check the KKT condition for the case when the optimal storage decision satisfies $-\underline{e} < e_t^\ast < 0$ with {\em nonbinding constraints}. The optimization in this case reduces to
\beq \label{eq:opt_CaseIII-b}
\begin{array}{ll}
\underset{(d_{tk}),e_t}{\mbox{maximize}} & \sum_k U_{tk}(d_{tk}) - \frac{\gamma}{\rho}[e_t]^-\\
\mbox{subject to} & \\
(\lambda): & \sum_k d_k  = g_t + [e_t]^-\\
(\underline{\mu}^-,\bar{\mu}^-): & 0 \le [e_t]^- \le \underline{e}\\
(\underline{\eta}_k,\bar{\eta}_k): & 0 \le d_{tk} \le \bar{d}_{tk}.\\
\end{array}
\eeq
The KKT condition requires that $\lambda^\ast=\gamma/\rho$, which gives
\[
d^\ast_{tk} = f_{tk}(\gamma/\rho),~~e_t^\ast=g_t-f_t(\gamma/\rho).
\]

\underline{Case III-c ($\sigma^{+o}_t \le g_t \le \sigma^{-o}_t$):}
 We check the KKT condition for the case when the optimal storage decision satisfies $e_t^\ast = 0$. The optimization in this case reduces to
\beq \label{eq:opt_CaseIII-c}
\begin{array}{ll}
\underset{(d_{tk})}{\mbox{maximize}} & \sum_k U_{tk}(d_k) \\
\mbox{subject to} & \\
(\lambda): & \sum_k d_{tk}  = g_t \\
(\underline{\eta}_k,\bar{\eta}_k): & 0 \le d_{tk} \le \bar{d}_{tk}.\\
\end{array}
\eeq
The KKT condition requires that
\[
d^\ast_{tk} = f_{tk}(\lambda^\ast)~~\Rightarrow~~\sum_k f_{tk}(\lambda^\ast)=g_t.
\]
We therefore have $d^\ast_{tk}=f_{tk}(f^{-1}_t(g_t))$ and $e^\ast_t=0$.

\underline{Case III-d ($\sigma^{-o}_t < g_t < \sigma^{-}_t$):}
The proof in this case is parallel to Case III-b.

 We check the KKT condition for the case when the optimal storage decision satisfies $0 < e_t^\ast < \overline{e}$ with {\em nonbinding constraints}. The optimization in this case reduces to
\beq \label{eq:opt_CaseIII-d}
\begin{array}{ll}
\underset{(d_{tk}),e_t}{\mbox{maximize}} & \sum_k U_{tk}(d_{tk}) + \gamma\tau [e_t]^+\\
\mbox{subject to} & \\
(\lambda): & \sum_k d_k  = g_t - [e_t]^+\\
(\underline{\mu}^+,\bar{\mu}^+): & 0 \le [e_t]^+ \le \overline{e}\\
(\underline{\eta}_k,\bar{\eta}_k): & 0 \le d_{tk} \le \bar{d}_{tk}.\\
\end{array}
\eeq
The KKT condition requires that $\lambda^\ast=\gamma\tau$, which gives
\[
d^\ast_{tk} = f_{tk}(\gamma\tau),~~e_t^\ast=g_t-f_t(\gamma\tau).
\]

\underline{Case III-e ($\sigma^{-}_t \le g_t < \Delta^-_t$):}
The proof in this case is parallel to Case III-a.

We check the KKT condition for the case that $e_t^\ast=\overline{e}$.  The optimization (\ref{eq:opt_zero}) reduces to
\beq \label{eq:opt_CaseIII-e}
\begin{array}{ll}
\underset{(d_{tk})}{\mbox{maximize}} & \sum_k U_{tk}(d_{tk})\\
\mbox{subject to} & \\
(\lambda): & \sum_k d_{tk}  = g_t - \overline{e}\\
(\underline{\eta}_k,\bar{\eta}_k): & 0 \le d_{tk} \le \bar{d}_{tk}\\
\end{array}
\eeq
from which we have $d_{tk}^\ast=f_{tk}(f_t^{-1}(g_t-\overline{e}))$.

Therefore in summary,
\bea
e^\ast_t(g_t)&=&\left\{\begin{array}{ll}
-\underline{e}, & g_t \le \sigma^+_t\\
g_t-\sigma_t^{+o}, & g_t\in (\sigma^+_t, \sigma^{+o}_t)\\
0, & g_t \in [\sigma^{+o}_t, \sigma^{-o}_t]\\
 g_t- \sigma_t^{-o}, & g_t \in (\sigma^{-o}_t, \sigma^-_t)\\
\overline{e}, &  g_t \ge \sigma^-_t,\\
\end{array}\right.\label{eq:eastNSL}\\
d^\ast_t(g_t)&=& z^\ast_t - e^\ast_t + g_t \label{eq:dastNSl}.\\
d_{tk}^\ast(g_t) &=& \left\{\begin{array}{ll}
f_{tk}(\pi^+_t), & g_t < \Delta^+_t\\
f_{tk}(f^{-1}_t(g_t+\underline{e})), & g_t \in [\Delta_{t}^+,\sigma_{t}^+]\\
f_{tk}(f^{-1}_t(\gamma/\rho)), & g_t\in (\sigma^+_t, \sigma^{+o}_t)\\
f_{tk}(f^{-1}_t(g_t)), &  g_t \in [\sigma^{+o}_t, \sigma^{-o}_t]\\
f_{tk}(f^{-1}_t(\tau \gamma)), & g_t\in (\sigma^{-o}_t,\sigma_t^-)\\
f_{tk}(f^{-1}_t(g_t-\overline{e})), & g_t\in [\sigma^{-}_t,\Delta_t^-]\\
f_{tk}(\pi^-_t), & g_t > \Delta_{t}^-.\\
\end{array}\right.\label{eq:dastkNSl}
\eea

Lastly, we can see that the optimal decisions that maximizes the stochastic dynamic program (\ref{eq:optimization}) under the sufficient optimality condition, \ie equations (\ref{eq:OptNetConsumptionNSL}), and (\ref{eq:eastNSL})--(\ref{eq:dastkNSl}), are equivalent to the MCO algorithm's schedule, hence the MCO algorithm is the optimal solution to the stochastic dynamic program if (i) $B > 2T\tau \overline{e}$, and (ii) $s \in(T \underline{e} / \rho, B-T \tau \overline{e})$. \QED


\subsection{Proof of Corollary \ref{corol:OptDec}}
The proof follows directly from Theorem \ref{thm:sufficient}. By replacing $\underline{e}_t^\dagger$ and $\overline{e}_t^\dagger$ in the MCO with $\underline{e}$ and $\overline{e}$, respectively, we get the optimal co-optimization schedules $e^\ast_t$ and $d^\ast_{tk}$. \QED


\subsection{Proof of Theorem \ref{thm:Myopic}}
\input{MCOproof_v7}


\subsection{Proof of Corollary \ref{corol:NetConsMyopic}}
The proof of the corollary follows directly from Theorem \ref{thm:Myopic}. For every $t\in [0,T-1]$, by plugging the MCO's total consumption $d^\dagger_t(g_t)$ and storage operation $e^\dagger_t(g_t)$ into $z^\dagger_t(g_t)= d^\dagger_t(g_t)+ e^\dagger_t(g_t) - g_t$, we have (\ref{eq:OptNetConsumption}), which also shows that $z^\dagger_t$ is a monotonically decreasing
piecewise-linear function of $g_t$. \QED
\subsection{Proposition \ref{prop:PassivePros} and Proof of Proposition \ref{prop:PassivePros}}
\input{PassiveProp_v7}

%% file: appendix_lemmas_v7.tex
We present here three lemmas that highlight the structural properties of the optimal prosumer decisions of $\Pc$ under the sufficient optimality conditions: (i) $B > 2T\tau \overline{e}$, and (ii) $s \in(T \underline{e} / \rho, B-T \tau \overline{e})$.
To this end, we introduce Bellman's optimality condition. For $t\in[0, T-1]$, the Bellman's optimality condition is given by
\begin{align}
    \label{eq:DERBellman1}
    V_t(x_t) &= \max_{u_t \in \mathcal{A}(x_t)} \Big[r_t(x_t,u_t)+\mathbb{E}\Big( V_{t+1}(x_{t+1})\Big|x_t\Big)\Big],
\end{align}
where $\Ac(x_t)$ is the constraint set on $u_t$ defined in $\Pc$.


\begin{lemma}[Marginal value of storage]\label{prop:MVoS}  The value function $V_t(\cdot)$ is monotonically increasing for all $t$.  For all $\Delta \in [-\underline{e}/\rho,\bar{e}\tau]$ and $s_t+\Delta \in [0, B]$,
\beq \label{eq:MVoS}
V_t(s_t+\Delta,g_t)=V_t(x_t)+\gamma\Delta~~\Rightarrow~~\frac{\partial}{\partial s}V_t(x_t) = \gamma,
\eeq
\ie  the marginal value of energy in the storage is $\gamma$.
\end{lemma}

\subsection*{Proof of Lemma~\ref{prop:MVoS}}
We show first that the value function $V_t(x_t)$ is monotonically increasing.
From Bellman's optimality condition (\ref{eq:DERBellman1}),
for all $\epsilon>0$,
\[
V_t(s_t + \epsilon, g_t) \ge \gamma \epsilon + V_t(s_t,g_t) >  V_t(s_t,g_t), \forall s_t \in (0,B),
\]
where the first inequality comes from a suboptimal policy for which the $\epsilon$-level of energy in the storage is not used in the decision process after $t$. Likewise,
\[
V_t(s_t,g_t+\epsilon) \ge  V_t(s_t,g_t) + \pi^-_t \epsilon >  V_t(s_t,g_t),
\]
where the first inequality comes from that the $\epsilon$ additional generation exported to the grid. Therefore, the value function $V_t$ is monotonically increasing with the system state $x_t$.

Next we show (\ref{eq:MVoS}) by backward induction. Let $\Delta_s=(\Delta,0)$. At $t=T-1$,
\begin{equation}
  \begin{aligned}
  &V_{T-1}(x_{T-1}+\Delta_s)=\max_{u=(\bm{d},e)} \bigg[r_{T-1}(x_{T-1}+\Delta_s,u)\\
  &\hspace{4em} +\gamma(s_{T-1}+\Delta+\tau[e]^+-[e]^-/\rho)\bigg]\\
    &~~\stackrel{(a)}{=}\max_{u=(\bm{d},e)} \bigg[r_{T-1}(x_{T-1},u)\\
   &\hspace{4em} +\gamma(s_{T-1}+\tau[e]^+-[e]^-/\rho)\bigg] + \gamma \Delta\\
    &~~=V_{T-1}(x_{T-1}) + \gamma\Delta, \nn
  \end{aligned}
\end{equation}
where (a) is from that the stage-reward $r_t(\cdot)$ in (\ref{eq:stageReward}) does not depend on $s_t$ unless $t=T$.

Assuming (\ref{eq:MVoS}) holds for $t+1$, we have, in interval $t$,
\begin{equation}
  \begin{aligned}[b]
&V_t(x_t+\Delta_s) =\max_{u=(\bm{d},e)}\bigg[r_{t}(x_{t}+\Delta_s,u)\\
&\hspace{4em} + \mbbE\Big(V_{t+1}(s_t+\tau[e]^+-[e]^-/\rho + \Delta,g_{t+1})|x_t\Big)\bigg]\\
&~~=\gamma \Delta+ \max_{u=(\bm{d},e)} \bigg
[r_t(x_t,u) \\
&\hspace{4em} + \mbbE\Big(V_{t+1}(s_t+\tau[e]^+-[e]^-/\rho,g_{t+1})\big|x_t\Big)\bigg]\\
&~~= V_t(x_t) + \gamma \Delta. \nn
\end{aligned}
\end{equation}
 \QED
 

\begin{lemma}[Storage-consumption complementarity]  \label{prop:complementarity} Under $B > 2T\tau \overline{e}$, and $s \in(T \underline{e} / \rho, B-T \tau \overline{e})$, and for all $t$, the optimal co-optimization decisions obey (a) $e^\ast_tz_t^\ast \le 0$, and (b) $e^\ast_t \tilde{d}^\ast_t \le 0$, where $\tilde{d}^\ast_t:= d_t^\ast - g_t$ is the DG-adjusted total consumption.
\end{lemma}

\subsection*{Proof of Lemma~\ref{prop:complementarity}}
\paragraph{Proof of $e^\ast_t z_t^\ast \le 0$}  Proof by contradiction.  Suppose that $e_t^\ast  > 0$, when $z^\ast_t>0$. Let $\tilde{e}_t :=  e^\ast_t-\epsilon >0$ for some $\epsilon>0$ such that $\tilde{z}_t:= d^\ast_t + \tilde{e}_t - g_t  > 0$.

Consider the stage reward $r_t(x_t,(\tilde{e}_t,\bm{d}_t^\ast))$ in interval $t$:
\bea
r_t(x_t,(\tilde{e}_t,\bm{d}_t^\ast)) &=& U_t(\bm{d}_t^\ast) - \pi^+_t(d_t^\ast+\tilde{e}_t - g_t) \nn\\
& =  & r_t(x_t,(e_t^\ast,\bm{d}_t^\ast)) + \pi^+_t \epsilon.\nn
\eea
The value function in interval $t$ is given by
\begin{equation}
  \begin{aligned}[b]
V_t(x_t)&= r_t(x_t,(e_t^\ast,\bm{d}_t^\ast))+ \mbbE\Big(V_{t+1}(s_t + \tau e^\ast_t, g_{t+1})\Big|x_t\Big)\\
 &=r_t(x_t,(\tilde{e}_t+\epsilon,\bm{d}_t^\ast))\\
 &~~~~+ \mbbE\Big(V_{t+1}\big(s_t + \tau\tilde{e}_t + \tau \epsilon, g_{t+1}\big)\Big|x_t\Big)\\
 &=r_t(x_t,(\tilde{e}_t,\bm{d}_t^\ast))+\mbbE\Big(V_{t+1}\big(s_t + \tau \tilde{e}_t, g_{t+1}\big)\Big|x_t\Big) \\
 &~~~~-\epsilon(\pi^+_t-\tau \gamma)\\
 &<r_t(x_t,(\tilde{e}_t,\bm{d}_t^\ast))+  \mbbE\Big(V_{t+1}\big(s_t + \tau \tilde{e}_t, g_{t+1}\big)\Big|x_t\Big),\nn
\end{aligned}
\end{equation}
where the last inequality uses the assumption in (\ref{eq:gamma}). Therefore, $e_t^\ast  > 0$ cannot be optimal.

 Next, suppose that $e_t^\ast  < 0$ and $z^\ast_t<0$. Let $\tilde{e}_t :=  e^\ast_t+\epsilon<0$ for some $\epsilon>0$ such that $d^\ast_t + \tilde{e}_t - g_t  < 0$.

Consider the stage reward $r_t(x_t,(\tilde{e}_t,\bm{d}_t^\ast))$ in interval $t$:
\bea
r_t(x_t,(\tilde{e}_t,\bm{d}_t^\ast)) &=& U(\bm{d}_t^\ast) - \pi^-_t(d_t^\ast+\tilde{e}_t - g_t) \nn\\
& =  & r_t(x_t,(e^\ast_t,\bm{d}_t^\ast)) - \pi^-_t \epsilon.\nn
\eea
The value  function in interval $t$ is given by
\begin{equation}
  \begin{aligned}[b]
V_t(x_t)&= r_t(x_t,(e_t^\ast,\bm{d}_t^\ast))+ \mbbE\Big(V_{t+1}\big(s_t + e^\ast_t/\rho, g_{t+1}\big)\Big|x_t\Big)\\
 &=r_t(x_t,(\tilde{e}_t-\epsilon,\bm{d}_t^\ast))\\
 &~~~~+ \mbbE\Big(V_{t+1}\big(s_t + \frac{1}{\rho}(\tilde{e}_t - \epsilon), g_{t+1}\big)\Big|x_t\Big)\\
 &=r_t(x_t,(\tilde{e}_t,\bm{d}_t^\ast)) +\mbbE\Big(V_{t+1}\big(s_t + \tilde{e}_t/\rho, g_{t+1}\big)\Big|x_t\Big) \\
 &~~~~- \epsilon(\gamma/\rho-\pi^-_t)\\
 &<r_t(x_t,(\tilde{e}_t,\bm{d}_t^\ast))+  \mbbE\Big(V_{t+1}\big(s_t + \tilde{e}_t/\rho, g_{t+1}\big)\Big|x_t\Big).\nn
\end{aligned}
\end{equation}
Therefore, $e_t^\ast  < 0$ cannot be optimal.

\paragraph{Proof of $e^\ast_t(d_t^\ast-g_t)\le 0$} If $z^\ast_t>0$, by the complementarity property, $e^\ast_t\le 0$. With $z^\ast_t =d^\ast_t+e^\ast_t-g_t>0$, we have $d^\ast_t-g_t >0$, hence $e^\ast_t(d_t^\ast-g_t)\le 0$.

Likewise,   If $z^\ast_t<0$, then  $e^\ast_t\ge 0$, implying $d^\ast_t-g_t < 0$.  Again,
$e^\ast_t(d_t^\ast-g_t)\le 0$. \QED


\begin{lemma}[Optimal storage operation]\label{prop:OptS}
 Under $B > 2T\tau \overline{e}$, and $s \in(T \underline{e} / \rho, B-T \tau \overline{e})$, and for all $t$, at the optimal total consumption $d_t^\ast$, we have
\beq \label{eq:east}
e^\ast_t = \left\{\begin{array}{ll}
\max\{g_t-d^\ast_t, -\underline{e}\}, & g_t \le d^\ast_t\\
\min\{g_t-d^\ast_t, \overline{e}\}, & g_t> d^\ast_t,\\
\end{array}
\right.
\eeq
and $e_t^\ast \le g_t$.
\end{lemma}

\subsection*{Proof of Lemma~\ref{prop:OptS}}
From the optimality equation with the optimal consumption $\bm{d}_t^\ast$ and total consumption $d^\ast_t={\bm 1}^{\top}\bm{d}^\ast_t$,
\bea
V_t(x)  &  = & \max_{e_t \in [-\underline{e},\overline{e}]} \Big[r_t(x_t,(e_t,\bm{d}_t^\ast))\nn \\& & + \mbbE\Big(V_{t+1}(s_t + \tau[e_t]^+-[e_t]^-/\rho, g_{t+1})\Big|x_t\Big)\Big]\nn\\
&=&\max_{e_t \in [-\underline{e},\overline{e}]} \Big[U_t(\bm{d}_t^\ast) - P^{\mbox{\tiny NEM}}_{\pi_t}(d^\ast_t+e_t-g_t) \nn\\& &+ \mbbE\Big(V_{t+1}(s_t, g_{t+1})\Big|x_t\Big) + \gamma \tau [e_t]^+ - \gamma [e_t]^-/\rho \Big]\nn\\
&=&\hspace{-1em} \max_{e_t \in [-\underline{e},\overline{e}]} \Big[(\tau\gamma-\pi_t(e_t))[e_t]^++(\pi_t(e_t)-\gamma/\rho)[e_t]^-\nn\\
& & \hspace{4em} - \pi_t(e)(d^\ast_t-g_t)\Big] + C_t ,\label{eq:Vt}
\eea
where $C_t$ is a constant independent of $e$ and
\bea
\pi_t(e_t) =\left\{\begin{array}{ll} \pi_t^+, & d^\ast_t+e_t-g_t \ge 0\\
\pi_t^-, & \mbox{otherwise.}
\end{array}\right.\nn
\eea

\underline{Case I ($d_t^\ast\ge g_t$):}  By the complementarity property, we have
$e_t^\ast\le 0$, and $z_t^\ast\ge 0$.  Let    the feasible regions of $e_t$  be $\Emsc^+_t:=[-\underline{e},\overline{e}] \cap  \{e_t: g_t-d_t^\ast \le e_t \le 0\}$, within which $[e_t]^+=0$ and $\pi_t(e_t)=\pi^+_t$.
The optimization in (\ref{eq:Vt}) becomes
\[
  \begin{aligned}[b]
\max_{e_t\in \Emsc^+_t} & \Big[(\tau \gamma-\pi_t(e_t))[e_t]^+  +(\pi_t(e_t)-\gamma/\rho)[e_t]^-\\
& ~~~~- \pi_t(e_t)(d^\ast_t-g_t)\Big]\nn\\
= &\max_{e_t\in \Emsc^+_t}  \Big[(\pi_t^+-\gamma/\rho)[e_t]^--\pi_t^+(d_t^\ast-g_t)\Big].\\
\end{aligned}
\]
With $\pi_t^+\ge \gamma/\rho$ and $e_t \ge g_t-d_t^\ast$, we have
\beq\label{eq:e*_t1}
e^\ast_t=\max\{-\underline{e}, g_t-d_t^\ast\}.
\eeq

\underline{Case II ($d_t^\ast \le  g_t$):} By the complementarity property, we have $e_t^\ast \ge  0$, and $z_t^\ast\le 0$.  Let the feasible regions of $e_t$  be $\Emsc^-_t:=[-\underline{e},\overline{e}] \cap  \{e_t: e_t \le  g_t-d_t^\ast\}$, within which $[e_t]^-=0$ and $\pi_t(e_t)=\pi^-_t$.
The optimization in (\ref{eq:Vt}) becomes
\[
  \begin{aligned}[b]
\max_{e_t\in \Emsc^-_t} & \Big[(\tau \gamma-\pi_t(e_t))[e_t]^+  +(\pi_t(e_t)-\gamma/\rho)[e_t]^-\\
& ~~~~- \pi_t(e_t)(d^\ast_t-g_t)\Big]\nn\\
= &\max_{e_t\in \Emsc^-_t}  \Big[(\tau\gamma - \pi_t^-)[e_t]^+-\pi_t^-(d_t^\ast-g_t)\Big]\\
\end{aligned}
\]
With $\pi_t^- \le \tau\gamma$ and $e_t \le g_t-d^\ast_t$, we have
\beq \label{eq:e*_t2}
e^\ast_t = \min\{\overline{e},g_t-d_t^\ast\}.
\eeq
Combining (\ref{eq:e*_t1}-\ref{eq:e*_t2}), we have (\ref{eq:east}).

To show $e_t^\ast\le g_t$, we note that $g_t\ge e_t^\ast$ when $e_t^\ast\le 0$.
When $e_t^\ast > 0$, $z_t^\ast = d_t^\ast+e_t^\ast - g_t \le 0$,  by the complementarity condition.  therefore, $d_t^\ast \ge 0$ implies $e_t^\ast\le g_t$.
\QED

%% file: MCOproof_v7.tex
Recall the myopic optimization $\mathcal{P}_t$ in (\ref{eq:Opt_myopic}) 
\begin{subequations}
\begin{align*}
   \Pc_t: \max_{\bm{d}_t \in \mathbb{R}^K_+,e_t \in \mathbb{R}} &~~ U_t(\bm{d}_t) - P_{\pi_t}^{\mbox{\tiny NEM}}(z_t) + \gamma (\tau [e_t]^+-\frac{1}{\rho}[e_t]^-)\nonumber\\
   \text { subject to:~~} & z_t = \bm{1}^\top \bm{d}_t+ [e_t]^+-[e_t]^- -g_t \\
    & 0 \leq [e_t]^- \leq \min\{\underline{e},\rho s_t\} \nonumber\\
    & 0 \leq [e_t]^+ \leq \min\{\overline{e},(B-s_t)/\tau\}\nonumber \\
    & \bm{0} \preceq \bm{d}_{t} \preceq \bm{\bar{d}} \nonumber,
\end{align*}
\end{subequations}
which can be equivalently reformulated to the following three convex optimizations for every $t$:
\bea
\mathcal{P}^{+}_t: & \underset{\bm{d}_t \in \mathbb{R}^K_+,e_t \in \mathbb{R}}{\text{minimize}} &  \hspace{-0.3cm}\pi^+_t \cdot z_t - U_t(\bm{d}_t) - \gamma (\tau [e_t]^+-\frac{1}{\rho}[e_t]^-)\nonumber \\
& \text { subject to:~~} & z_t = \bm{1}^\top \bm{d}_t+ [e_t]^+-[e_t]^- -g_t \geq 0\nonumber \\
    && 0 \leq [e_t]^- \leq \min\{\underline{e},\rho s_t\} \nonumber\\
    && 0 \leq [e_t]^+ \leq \min\{\overline{e},(B-s_t)/\tau\}\nonumber\\
    && \bm{0} \preceq \bm{d}_{t} \preceq \bm{\bar{d}}\nonumber\\
\mathcal{P}^{-}_t: & \underset{\bm{d}_t \in \mathbb{R}^K_+,e_t \in \mathbb{R}}{\text{minimize}} &  \hspace{-0.3cm}\pi^-_t \cdot z_t - U_t(\bm{d}_t) - \gamma (\tau [e_t]^+-\frac{1}{\rho}[e_t]^-)\nonumber \\
& \text { subject to:~~} & z_t = \bm{1}^\top \bm{d}_t+ [e_t]^+-[e_t]^- -g_t \leq 0\nonumber \\
    && 0 \leq [e_t]^- \leq \min\{\underline{e},\rho s_t\} \nonumber\\
    && 0 \leq [e_t]^+ \leq \min\{\overline{e},(B-s_t)/\tau\}\nonumber\\
    && \bm{0} \preceq \bm{d}_{t} \preceq \bm{\bar{d}}\nonumber\\
\mathcal{P}^{0}_t: & \underset{\bm{d}_t \in \mathbb{R}^K_+,e_t \in \mathbb{R}}{\text{minimize}} & - U_t(\bm{d}_t) - \gamma (\tau [e_t]^+-\frac{1}{\rho}[e_t]^-)\nonumber \\
& \text { subject to:~~} & z_t = \bm{1}^\top \bm{d}_t+ [e_t]^+-[e_t]^- -g_t =0\nonumber \\
    && 0 \leq [e_t]^- \leq \min\{\underline{e},\rho s_t\}\nonumber \\
    && 0 \leq [e_t]^+ \leq \min\{\overline{e},(B-s_t)/\tau\}\nonumber\\
    && \bm{0} \preceq \bm{d}_{t} \preceq \bm{\bar{d}}\nonumber
\eea
Given $g_t$, the optimal schedule is the one that achieves the minimum value among $\mathcal{P}^{+}_t, \mathcal{P}^{-}_t$ and $\mathcal{P}^{0}_t$. Note that, for all three optimizations, the optimal consumption and storage operation exist. Because the Slater’s condition is satisfied for these optimizations, KKT conditions for optimality is necessary and sufficient.

Lemmas \ref{Lem:MyopicThm}--\ref{Lem:MyopicThm0} next are used to prove Theorem \ref{thm:Myopic}.

\begin{lemma}[Co-optimized myopic DER schedules under $\mathcal{P}^{+}_t$ and $\mathcal{P}^{-}_t$]\label{Lem:MyopicThm}
    For every time $t$ and device $k$ under the myopic co-optimization $\mathcal{P}_t$, it is optimal to consume ($f_{tk}(\pi^+_t)$) and discharge the storage at ($\min\{\underline{e},\rho s_t\}=: \underline{e}^\dagger_t$) when $g_t < \Delta_t^{+}:= \max\{f_t(\pi^+_t)-\underline{e}^\dagger_t,0\}$, and consume ($f_{tk}(\pi^-_t)$) and charge the storage at ($\min\{\overline{e},(B-s_t)/\tau\}=:\overline{e}^\dagger_t$), when $g_t > \Delta_t^{-}:= \max\{f_t(\pi^-_t)+\overline{e}^\dagger_t,0\}$.
\end{lemma}

\subsection*{Proof of Lemma~\ref{Lem:MyopicThm}}
First, we prove the optimal consumption and storage schedules when $g_t < \Delta_t^{+}:= \max\{f_t(\pi^+_t)-\underline{e}^\dagger_t,0\}$, leading to $z^\dagger_t>0$, are $d^\dagger_{tk} =  f_{tk}(\pi^+_t), e^\dagger_{t}= -\min\{\underline{e},\rho s_t\}$, respectively.

Given that the third term in the objective of $\mathcal{P}^{+}_t$ is independent from $\bm{d}_t$, the optimal consumption is equivalent to the no-storage case in \cite{Alahmed&Tong:22TSG}, which gives $d^\dagger_{tk}= f_{tk}(\pi^+_t), \forall k$ and $d^\dagger_{t}= f_{t}(\pi^+_t)$. 

Now, we resolve $\mathcal{P}^{+}_t$ with storage operation as the only variable:
\bea
& \underset{e_t}{\text{minimize}} &  (\pi^+_t -\gamma \tau) [e_t]^+ -(\pi^+_t -\frac{\gamma}{\rho})[e_t]^-\nonumber \\
& \text { subject to:~~} & z_t = f_{t}(\pi^+_t)+ [e_t]^+-[e_t]^- -g_t \geq 0\nonumber \\
    && 0 \leq [e_t]^- \leq \min\{\underline{e},\rho s_t\}\nonumber \\
    && 0 \leq [e_t]^+ \leq \min\{\overline{e},(B-s_t)/\tau\},\nonumber
\eea
which yields $e^\dagger_t = -\min\{\underline{e},\rho s_t\}$. Note that $z^\dagger_t:= d^\dagger_{t} +e^\dagger_t - g_t>0$ because $g_t < \Delta_t^{+}$.

Similarly for $g_t > \Delta_t^{-}:= \max\{f_t(\pi^-_t)+\overline{e}^\dagger_t,0\}$, the optimal consumption from \cite{Alahmed&Tong:22TSG} is $d^\dagger_{tk} =  f_{tk}(\pi^-_t)$ (and $d^\dagger_{t}= f_{t}(\pi^-_t)$) , which is plugged in $\mathcal{P}^{-}_t$ to solve for the optimal storage operation
\bea
& \underset{e_t}{\text{minimize}} &  (\pi^-_t -\gamma \tau) [e_t]^+ -(\pi^-_t -\frac{\gamma}{\rho})[e_t]^-\nonumber \\
& \text { subject to:~~} & z_t = f_{t}(\pi^-_t)+ [e_t]^+-[e_t]^- -g_t \leq 0\nonumber \\
    && 0 \leq [e_t]^- \leq \min\{\underline{e},\rho s_t\}\nonumber \\
    && 0 \leq [e_t]^+ \leq \min\{\overline{e},(B-s_t)/\tau\},\nonumber
\eea
yielding $e^\dagger_t = \min\{\overline{e},(B-s_t)/\tau\}$. Note that $z^\dagger_t:= d^\dagger_{t} +e^\dagger_t - g_t<0$ because $g_t > \Delta_t^{-}$. \QED

\begin{lemma}[Co-optimized myopic DER schedules under $\mathcal{P}^{0}_t$]\label{Lem:MyopicThm0}
    For every time $t$ and device $k$ under the myopic co-optimization $\mathcal{P}_t$, when $g_t \in [\Delta^{+}_t,\Delta^{-}_t]$ it is optimal to consume as
\begin{equation*}
d_{tk}^\ast = \left\{\begin{array}{ll}
f_{tk}(f^{-1}_t(g_t+\underline{e}_t^\dagger)), & g_t \in [\Delta^{+}_t,\sigma_{t}^{+}]\\
f_{tk}(f^{-1}_t(\gamma/\rho)), & g_t\in (\sigma^{+}_t, \sigma^{+o}_t)\\
f_{tk}(f^{-1}_t(g_t)), &  g_t \in [\sigma^{+o}_t, \sigma^{-o}_t]\\
f_{tk}(f^{-1}_t(\tau \gamma)), & g_t\in (\sigma^{-o}_t,\sigma_t^{-})\\
f_{tk}(f^{-1}_t(g_t-\overline{e}_t^\dagger)), & g_t\in [\sigma^{-}_t,\Delta_t^{-}],
\end{array}\right.
\end{equation*}
    and to operate the storage as
    \begin{equation*}
e^\ast_t=\left\{\begin{array}{ll}
-\underline{e}^\dagger_t, & g_t \in [\Delta^{+}_t,\sigma_{t}^{+}]\\
g_t-\sigma_t^{+o}, & g_t\in (\sigma_{t}^{+}, \sigma_{t}^{+o})\\
0, & g_t \in [\sigma_{t}^{+o}, \sigma_{t}^{-o}]\\
 g_t- \sigma_t^{-o}, & g_t \in (\sigma_{t}^{-o}, \sigma_{t}^{-})\\
\overline{e}_t^\dagger, &  g_t \in [\sigma^{-}_t,\Delta_t^{-}].
\end{array}\right.
\end{equation*}
\end{lemma}

\subsection*{Proof of Lemma~\ref{Lem:MyopicThm0}}
The proof follows from Case III in Theorem \ref{thm:sufficient}, but with replacing $\overline{e}$ and $\underline{e}$ with $\overline{e}_t^\dagger$ and $\underline{e}_t^\dagger$, respectively. \qed

Combining the optimal solutions in Lemmas \ref{Lem:MyopicThm}-\ref{Lem:MyopicThm0} solves the myopic co-optimization $\mathcal{P}_t$ in (\ref{eq:Opt_myopic}). \qed

%% file: PassiveProp_v7.tex
\begin{proposition}[Optimal policy of passive prosumers \cite{AlahmedTong:22StorageTPEC}]\label{prop:PassivePros}
Under $B > 2T\tau \overline{e}$, and $s \in(T \underline{e} / \rho, B-T \tau \overline{e})$ and for any DG-passive feasible consumption bundle $\hat{\bm{d}}_t \in \mathcal{D}$, the optimal storage operation is to discharge/charge as much as possible to minimize the absolute value of net consumption:
\begin{align}\label{eq:PassiveOptimization}
\mu^\ast_t \in \underset{e_t \in \{-\underline{e},\bar{e}\}}{\text{arg min}} &~~ |z_t|.
\end{align}
for every $t=0,\ldots, T-1$,
\end{proposition}

\subsection*{Proof of Proposition \ref{prop:PassivePros}}
When the consumption is dropped from the set of decision variables, the MDP {\em policy} $\mu := (\mu_0,\ldots,\mu_{T-1})$ becomes a sequence of decision rules, $x_t \stackrel{\mu_t}{\rightarrow} u_t := e_t$, for all $x_t$ and $t$, that specifies storage operation in each interval. The optimal storage operation of the prosumer decision problem in (\ref{eq:optimization}), under the sufficient optimality condition, follows from Lemma \ref{prop:OptS}, with fixing the consumption at $\hat{d}_{tk}$ for all $t=0,\ldots,T-1$ and $k$ and the total consumption at $\hat{d}_{t}=\sum_k \hat{d}_{tk}$. Therefore, using Lemma \ref{prop:OptS}, we have:
\beq 
e^\ast_t(g_t) = \left\{\begin{array}{ll}
\max\{g_t-\hat{d}_{t}, -\underline{e}\} & g_t \le \hat{d}_{t}\\
\min\{g_t-\hat{d}_{t}, \bar{e}\} & g_t> \hat{d}_{t}.\\
\end{array}
\right.\nn
\eeq
Given
\bea
\max (a, b) \pm c &= \max (a \pm c, b \pm c )\nn\\
\min (a, b) \pm c &= \min (a \pm c, b \pm c),\nn
\eea
the optimal net-consumption $z^\ast_t(g_t):=\hat{d}_{t}-g_t+e^\ast_t$, can be written as:
\beq 
z^\ast_t(g_t) = \left\{\begin{array}{ll}
\max\{0, \hat{d}_{t}-g_t-\underline{e}\}, & z^\ast_t(g_t) \geq 0\\
\min\{0, \hat{d}_{t}-g_t+\bar{e}\}, & z^\ast_t(g_t) < 0,\\
\end{array}
\right.\nn
\eeq
which is simply
\beq \label{eq:PropOptNetCons}
z^\ast_t(g_t) = \left\{\begin{array}{ll}
\hat{d}_{t}-g_t-\underline{e}, & z^\ast_t(g_t) > 0\\
0, & z^\ast_t(g_t) =0\\
\hat{d}_{t}-g_t+\bar{e}, & z^\ast_t(g_t) < 0.\\
\end{array}
\right.
\eeq
Note that to solve (\ref{eq:PassiveOptimization}), for every $t=0,\ldots,T-1$, one can break:
\bea
\mathcal{P}: & \underset{e \in \{-\underline{e},\bar{e}\}}{\text{minimize}} & |z_t|,\nonumber
\eea
to the following three convex optimizations $\mathcal{P}^+, \mathcal{P}^-$ and $\mathcal{P}^0$:
\bea
\mathcal{P}^+: & \underset{e_t \in \{-\underline{e},\bar{e}\}}{\text{minimize}} & z_t \nn\\
& \text{subject to} & \hat{d}_{t}-g_t + e_t \geq 0 \nonumber\\
\mathcal{P}^-: & \underset{e_t \in \{-\underline{e},\bar{e}\}}{\text{minimize}} & -z_t \nn\\
& \text{subject to} & \hat{d}_{t}-g_t + e_t \leq 0 \nonumber\\
\mathcal{P}^0: & \underset{e_t \in \{-\underline{e},\bar{e}\}}{\text{minimize}} & z_t \nn\\
& \text{subject to} & \hat{d}_{t}-g_t + e_t = 0 \nonumber
\eea

Given $g_t$, the optimal schedule is the one that achieves the minimum value among $\mathcal{P}^+, \mathcal{P}^-$ and $\mathcal{P}^0$. Note that, for all three optimizations, the optimal storage operation exists. Because the Slater’s condition is satisfied for these optimizations, KKT conditions for optimality is necessary and sufficient.
\par From Theorem 1 in \cite{Alahmed&Tong:22TSG}, we can use the renewable DG adjusted consumption $\tilde{d}_t(g_t):= \hat{d}_{t}-g_t$ here to characterize the optimal storage operation, as:
\begin{equation*}
e^\ast_t(g_t) = \left\{\begin{array}{ll}
-\underline{e}, & -\tilde{d}_t(g_t)< -\underline{e} \\
-\tilde{d}_t(g_t), & -\tilde{d}_t(g_t) \in [-\underline{e},\overline{e}]\\
\overline{e}, & -\tilde{d}_t(g_t)> \overline{e}.\\
\end{array}
\right.
\end{equation*}
Using the optimal storage operation $e^\ast_t(g_t)$ the optimal net consumption is
\bea
z^\ast_t(g_t) &=& \tilde{d}_t(g_t)+e^\ast_t(g_t)= \left\{\begin{array}{ll}
\tilde{d}_t(g_t)-\underline{e}, & z^\ast_t(g_t)>0 \\
0, & z^\ast_t(g_t)=0\\
\tilde{d}_t(g_t)+\overline{e}, & z^\ast_t(g_t)<0,
\end{array}
\right.\nonumber
\eea
which is equivalent to (\ref{eq:PropOptNetCons}). \QED

%% file: AppendixB_v7.tex
Here, we consider cases when the salvage value assumption in (\ref{eq:gamma}) does not hold, for all $t=0,\ldots,T-1$. We note that if (\ref{eq:gamma}) does not hold, we have the following cases, for $t=0,\ldots,T-1$,
\begin{enumerate}
    \item $\tau \gamma, \gamma/\rho \notin [\max \{(\pi^-_t)\},\min\{(\pi^+_t)\}]$.
    \begin{enumerate}[a)]
        \item $\tau \gamma< \gamma/\rho < \max \{(\pi^-_t)\} \leq \min \{(\pi^+_t)\}$.
        \item $\gamma/\rho > \tau \gamma > \min\{(\pi^+_t)\} \geq \max \{(\pi^-_t)\}$.
        \item $\gamma/\rho >\min\{(\pi^+_t)\} >\max \{(\pi^-_t)\} >\tau \gamma$.
    \end{enumerate}
    \item $\min \{(\pi^+_t)\} > \gamma/\rho > \max \{(\pi^-_t)\} > \tau \gamma$.
    \item $\gamma/\rho >\min \{(\pi^+_t)\} > \tau \gamma > \max \{(\pi^-_t)\}$.
\end{enumerate}
The first case above is when both of the efficiency-adjusted salvage values lie outside of $[\max \{(\pi^-_t)\},\min\{(\pi^+_t)\}]$. The second and third cases are when one of the efficiency-adjusted salvage values lies outside, while the other lies inside. We formalize the optimal co-optimization policy under case 1--3 in Propositions \ref{prop:PolicyCase1}--\ref{prop:PolicyCase3}, respectively.\footnote{The proofs of the three propositions follow the same idea in the proof of Theorem \ref{thm:sufficient}, but they are shorter given that the policies involve less thresholds.}

\subsection*{Optimal Policies under Case 1}

\begin{proposition}[Optimal policy under case 1]\label{prop:PolicyCase1}
Under $B > 2T\tau \overline{e}$, and $s \in(T \underline{e} / \rho, B-T \tau \overline{e})$, and for all $t$, the optimal policy, when a) $\gamma/\rho < \max \{(\pi^-_t)\} \leq \min \{(\pi^+_t)\}$ is to always maximally discharge the storage ($e^\ast_t = -\underline{e}$), whereas if b) $\tau \gamma > \min\{\pi^+_t\} \geq \max \{(\pi^-_t)\}$ it is optimal to always maximally charge the storage ($e^\ast_t =  \bar{e}$). If c) $\gamma/\rho >\min\{\pi^+_t\} >\max \{(\pi^-_t)\} >\tau \gamma$ it is optimal to keep the storage idle ($e^\ast_t =0$).
\par The consumption in all three cases (a)--(c) is a monotonically increasing function of $g_t$ and abides by two renewable-generation-independent thresholds as
    \begin{equation*}
    d^\ast_{tk}(g_t)=\begin{cases}
f_{tk}(\pi^+_t), &  g_t\leq  f_{t}(\pi^+_t)+e^\ast_t \\ 
f_{tk}\left(f^{-1}_t\left( g_t-e^\ast_t\right)\right), &\hspace{-0.3cm}  f_{t}(\pi^+_t)+e^\ast_t \leq g_t \leq f_{t}(\pi^-_t)+e^\ast_t \\ 
f_{tk}(\pi^-_t), &  g_t \geq f_{t}(\pi^-_t) + e^\ast_t.
\end{cases}
\end{equation*}
The net-consumption in all three cases is therefore a monotonically decreasing function of $g_t$:
\begin{equation*}
    z^\ast_t(g_t) = \begin{cases}
f_{t}(\pi^+_t) + e^\ast_t - g_t, &  g_t\leq  f_{t}(\pi^+_t)+e^\ast_t \\ 
0, &\hspace{-0.2cm}  f_{t}(\pi^+_t)+e^\ast_t \leq g_t \leq f_{t}(\pi^-_t)+e^\ast_t \\ 
f_{t}(\pi^-_t) + e^\ast_t - g_t, &  g_t \geq f_{t}(\pi^-_t) + e^\ast_t.
\end{cases}
\end{equation*}

\end{proposition}
\subsection*{Optimal Policy under Case 2}
\begin{proposition}[Optimal policy under case 2]\label{prop:PolicyCase2}
Under $B > 2T\tau \overline{e}$, and $s \in(T \underline{e} / \rho, B-T \tau \overline{e})$ and for all $t$, if $\min \{(\pi^+_t)\} > \gamma/\rho > \max \{(\pi^-_t)\} > \tau \gamma$, the optimal co-optimization policy schedules the storage as an increasing function of $g_t$:
\begin{align*}
    e^\ast_{t}(g_t)= \begin{cases}
-\underline{e}, &  g_t\leq  \sigma^+_{t} \\
g_t-\sigma^{+o}_{t}, &  \sigma^+_{t} \leq g_t \leq \sigma^{+o}_{t} \\ 
0, &  g_t > \sigma^{+o}_{t}, \\ 
\end{cases}
\end{align*}
and the consumption also as an increasing function of $g_t$:
     \begin{align*}
    d^\ast_{tk}(g_t) = \begin{cases}
f_{tk}(\pi^+_t), &  g_t\leq  \Delta^+_{t} \\
f_{tk}\left(f^{-1}_t\left( g_t+\underline{e}\right)\right), &  \Delta^+_{t} \leq g_t\leq  \sigma^+_{t} \\
f_{tk}(\gamma/\rho), &  \sigma^+_{t} \leq g_t \leq \sigma^{+o}_{t} \\ 
f_{tk}\left(f^{-1}_t\left( g_t\right)\right), &  \sigma^{+o}_{t} \leq g_t \leq f_{t}(\pi^-_t) \\ 
f_{tk}(\pi^-_t), &  g_t \geq f_{t}(\pi^-_t).
\end{cases}
\end{align*}
This yields a net-consumption that is a monotonically decreasing function of $g_t$:
\begin{equation*}
    z^\ast_t(g_t) = \begin{cases}
f_{t}(\pi^+_t) -\underline{e} - g_t, &  g_t\leq  \Delta^+_{t} \\ 
0, &  \Delta^+_{t} \leq g_t \leq f_{t}(\pi^-_t) \\ 
f_{t}(\pi^-_t) - g_t, &  g_t \geq f_{t}(\pi^-_t) + e^\ast_t.
\end{cases}
\end{equation*}

\end{proposition}
\subsection*{Optimal Policy under Case 3}

\begin{proposition}[Optimal policy under case 3]\label{prop:PolicyCase3}
Under $B > 2T\tau \overline{e}$, and $s \in(T \underline{e} / \rho, B-T \tau \overline{e})$, and for all $t$, if $\gamma/\rho >\min \{(\pi^+_t)\} > \tau \gamma > \max \{(\pi^-_t)\}$, the optimal co-optimization policy schedules the storage as an increasing function of $g_t$:
 \begin{align*}
    e^\ast_{t}(g_t) = \begin{cases}
0, &  g_t < \sigma^{-o}_t \\ 
g_t-\sigma^{-o}_t, &  \sigma^{-o}_t \leq g_t \leq \sigma^{-}_t \\ 
\overline{e}, &  g_t \geq \sigma^{-}_t,
\end{cases}
\end{align*}
and the consumption also as an increasing function of $g_t$:
\begin{align*}
    d^\ast_{tk}(g_t) = \begin{cases}
f_{tk}(\pi^+_t), &  g_t\leq  f_{t}(\pi^+_t) \\ 
f_{tk}\left(f^{-1}_t\left( g_t\right)\right), &  f_{t}(\pi^+_t) \leq g_t \leq \sigma^{-o}_t \\ 
f_{tk}(\tau \gamma), &  \sigma^{-o}_t \leq g_t \leq \sigma^{-}_t \\ 
f_{tk}\left(f^{-1}_t\left( g_t-\overline{e}\right)\right), &  \sigma^{-}_t \leq g_t \leq \Delta^{-}_t \\ 
f_{tk}(\pi^-_t), &  g_t \geq \Delta^{-}_t.
\end{cases}
\end{align*}
This yields a net-consumption that is monotonically decreasing with $g_t$ as:
\begin{equation*}
    z^\ast_t(g_t) = \begin{cases}
f_{tk}(\pi^+_t) - g_t, &  g_t\leq  f_{t}(\pi^+_t) \\ 
0, &  f_{t}(\pi^+_t) \leq g_t \leq \Delta^{-}_t \\ 
f_{t}(\pi^-_t)+\overline{e} - g_t, &  g_t \geq \Delta^{-}_t.
\end{cases}
\end{equation*}

\end{proposition}

\subsection{Discussion}

In case (1), when $\tau \gamma > \min \{(\pi^+_t)\} \geq \max \{(\pi^-_t)\}$, the prosumer always charges the storage because the value of storing energy in the storage is higher than the value of discharging and achieving higher self-consumption ($\tau \gamma> \min \{(\pi^+_t)\}$). When $\gamma/\rho < \max \{(\pi^-_t)\}\leq \min \{(\pi^+_t)\}$, the prosumer always discharges the storage because the value of storing energy in the storage is below the value of exporting the energy to the grid ($\gamma/\rho < \pi^-_t, \forall t$). Note that unlike (\ref{eq:gamma})'s case of $\max \{(\pi^-_t)\}< \tau \gamma \leq \gamma/\rho < \min \{(\pi^+_t)\}$, the storage, for every $t$, charges from the grid when $\tau \gamma > \min \{(\pi^+_t)\} \geq \max \{(\pi^-_t)\}$, and discharges into the grid when $\gamma/\rho < \max \{(\pi^-_t)\}\leq \min \{(\pi^+_t)\}$. When $\gamma/\rho >\min \{(\pi^+_t)\} >\max \{(\pi^-_t)\} >\tau \gamma$, for every $t$, net exporting the DER is worth higher than charging storage. Also discharging the storage incurs a very high cost, therefore it is optimal neither to charge nor to discharge.
\par The net-consumption zone length is the shortest and the net-production zone length is the longest when the storage always discharges. The net-consumption zone length is the longest and the net-production zone length is the shortest when the storage always charges. The net-zero zone length is the same in all cases, which is $f_{t}(\pi^-_t) - f_{t}(\pi^+_t)$.
\par Case (2) shows that when the value of charging is small $\tau \gamma <\max \{(\pi^-_t)\}$ it is optimal to never charge the storage. The battery, therefore, discharges when the renewable output is small and stays idle when the renewable is high. The consumption of every device $k$ monotonically increases from $f_{tk}(\pi^+_t)$ to $f_{tk}(\pi^-_t)$ as $g_t$ increases.
\par Case (3) shows that when the cost of discharging is very high $\gamma/\rho >\min \{(\pi^+_t)\}$ it is optimal to never discharge the storage. The battery, therefore, charges when the renewable output is high and stays idle when the renewable is low. The consumption  of every device $k$ again monotonically increases from $f_{tk}(\pi^+_t)$ to $f_{tk}(\pi^-_t)$ as $g_t$ increases, but with modified thresholds and net-zero zone consumptions.

%% file: AppendixD_v7.tex
We incorporate storage degradation cost, which is the cost associated with charging and discharging repeatedly causing storage aging, by adopting a linear degradation cost model \cite{Shi&Xu&Wang&Zhang:18TPS,Chis&Koivunen:19TSG}. The battery cell replacement price $\chi_B$ and the number of battery cycles $N$ when its operation is limited by $s_t\in [\underline{s},\overline{s}], \forall t\in[0,T]$ with $\underline{s},\overline{s}$ being the minimum and maximum limits as percentage of the total battery capacity $B$, are used to linearize the battery degradation price $\chi$ as
\begin{equation}
    \chi = \frac{\chi_B}{2N (\overline{s}-\underline{s})}.
\end{equation}
Using the linearized battery degradation price $\chi \geq 0$, the storage degradation cost of charging/discharging one unit of energy is given, for every $ t\in[0,T-1]$, by
\begin{equation}\label{eq:DegradeCost}
    C_t(e_t)= \chi \cdot ([e_t]^++[e_t]^-).
\end{equation}
We next present how the degradation cost can be incorporated and how the program $\mathcal{P}$ can be re-formulated to show that the analysis in Sec.\ref{sec:prosumer} generalizes.

First we note that, from $s_{t}=s+ \sum_{i=0}^{t-1} \tau [e_i]^+-[e_i]^-/\rho$, we have $s_T- s = \sum_{t=0}^{T-1} (\tau [e_t]^+-[e_t]^-/\rho)$. Hence, the objective function without storage degradation cost can be written as
\begin{equation}\label{eq:re-modify}
\mathbb{E}_{\mu}\left\{\sum_{t=0}^{T-1} (\gamma \tau [e_t]^+- \frac{\gamma}{\rho}  [e_t]^-)+\sum_{t=0}^{T-1}   S_{\pi_t}^{\mbox{\tiny NEM}}(u_t;g_t) \right\},
\end{equation}
which we offer an analytical solution for under $B > 2T\tau \overline{e}$, and $s \in(T \underline{e} / \rho, B-T \tau \overline{e})$, and the assumption in (\ref{eq:gamma}),
$$\max\{(\pi_t^-)\} \le \tau \gamma \le \frac{\gamma}{\rho}  \le \min\{(\pi_t^+)\}.$$

Given the storage degradation cost, the reward function in (\ref{eq:stageReward}) is redefined to
\begin{equation}
        r_{t}\left(x_t,u_t\right) := \begin{cases} S_{\pi_t}^{\mbox{\tiny NEM}}(u_t;g_t) - C_t(e_t), & t \in[0, T-1] \\ \gamma (s_{T}-s), & t=T,\end{cases} \\
\end{equation}
and the objective of the storage-consumption co-optimization $\mathcal{P}$ becomes 
$$\mathbb{E}_{\mu}\left\{\gamma (s_{T}-s)+\sum_{t=0}^{T-1}   S_{\pi_t}^{\mbox{\tiny NEM}}(u_t;g_t) - C_t(e_t)\right\}.$$

Using $s_T- s = \sum_{t=0}^{T-1} (\tau [e_t]^+-[e_t]^-/\rho)$, the objective reformulates to
$$\mathbb{E}_{\mu}\left\{\sum_{t=0}^{T-1} (\gamma \tau-\chi) [e_t]^+- (\frac{\gamma}{\rho}+\chi)  [e_t]^-)+\sum_{t=0}^{T-1}   S_{\pi_t}^{\mbox{\tiny NEM}}(u_t;g_t) \right\},$$
which is effectively a minor modification to the no-storage-degradation case in (\ref{eq:re-modify}). Therefore, the storage degradation effect can be incorporated into the storage salvage value term, and the result is that instead of gaining a salvage value at $\gamma \tau$ when we charge under the no-degradation case, the gain reduces to ($\gamma \tau-\chi$) when degradation is incorporated. Similarly, instead of losing a salvage value at $\frac{\gamma}{\rho}$ when we discharge under the no-degradation case, the loss rate increases to ($\frac{\gamma}{\rho}+\chi$) when degradation is incorporated. Therefore, by modifying the salvage value assumption (\ref{eq:gamma}) to
$$\max\{(\pi_t^-)\} \le \gamma \tau-\chi \le \frac{\gamma}{\rho}+\chi  \le \min\{(\pi_t^+)\},$$
the analysis follows directly.

%% file: AppendixC_v7.tex
We show here additional results on how the benefits of co-optimization to prosumers and utility revenue recovery model. All consumption, renewables, storage, and tariff settings are the as in Sec.~\ref{sec:num}.

\subsection{Prosumer gain}
\begin{figure}
    \centering
    \includegraphics[scale=0.37]{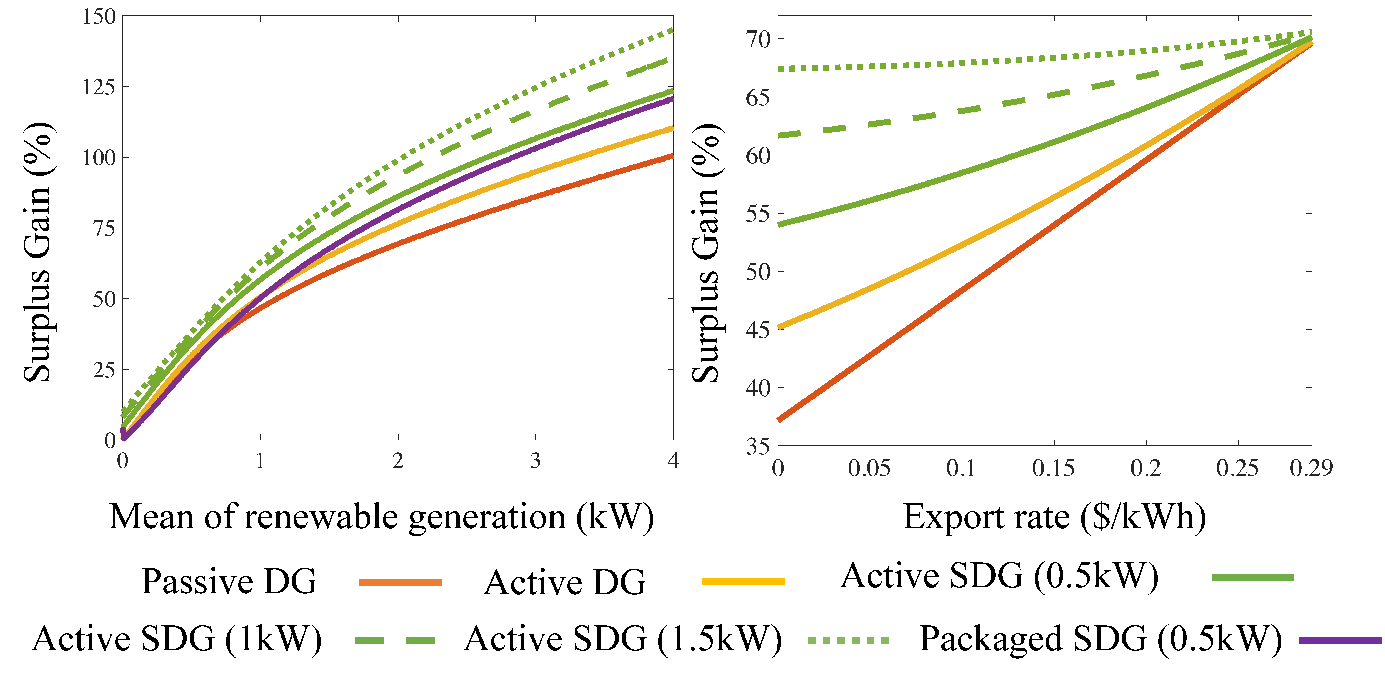}
    \vspace{-0.8cm}
    \caption{Cumulative surplus gain (\%) over consumer's surplus.}
    \label{fig:RewardGain}
\end{figure}

Fig.\ref{fig:RewardGain} shows the percentage gain of total prosumer surplus over that of a consumer with varying renewable generation (left) and export rate (right). As the renewable generation increased, prosumer surpluses gain increased, with SDG customers benefiting the most. Active SDG prosumers had the highest surplus gain for all $g_t$, which was also increasing as the storage charging/discharging powers increase from 0.5kW (solid) to 1.5 (dotted green). As the renewable generation increased, active and passive DG prosumers had the same surplus gains up to certain thresholds (0.75kW for DG customers) after which the curves separated, as the effect of active consumption took place. Note the non-monotonicity of the packaged SDG prosumer, as when the renewable generation is non-zero, the storage shifts from discharging to charging, regardless of the household's net consumption. On the other hand, the right panel of Fig.\ref{fig:RewardGain} shows the performance of active SDG prosumers, particularly under low export rates. The trajectories of the active SDG prosumers' surplus gains show how installing storage (particularly with higher charging/discharging powers) considerably absorbed the impact of reducing export rates. At \$0.05/kWh and with 0.5kW charging/discharging powers, active SDG prosumers achieved a 57\% surplus gain, whereas active DG and passive DG achieved 48\% and 41\%, respectively. The surplus differences diminished when the export rate approached the retail rate, since in NEM 1.0 ($\pi^+=\pi^-$), the grid can be used as a virtual storage, and the prosumer has no incentive to exercise DER-aware consumption decisions. 

\subsection{DSO Benefits}\label{subsec:DSObenefits}
Consumption-storage co-optimization reduces the power flow from net consumption and benefits the DSO indirectly. We capture some of these benefits by showing the effects of consumption-storage co-optimization on RPF that, if not well-regulated, can cause voltage instabilities and line losses. 

\par
\begin{figure}
    \centering
    \includegraphics[scale=0.35]{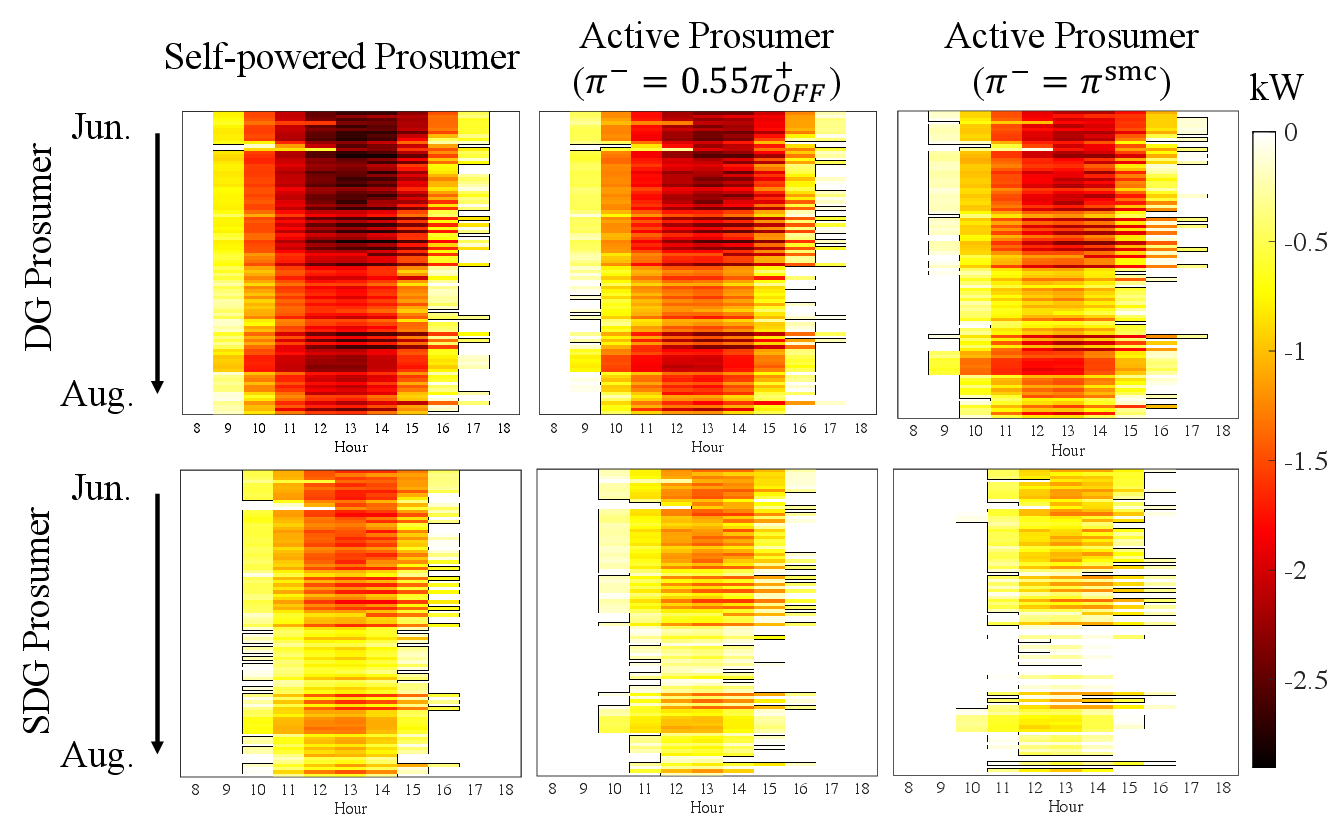}
    \vspace{-0.3cm}
    \caption{Hourly RPF (in kW) of different customer types ($\overline{e}=\underline{e}=1$kW).}
    \label{fig:RPF}
\end{figure}

 To evaluate the performance of the co-optimization policy in reducing RPF, we record the time instants and levels of RPF at the DSO revenue meter of the studied household under different customer types and export rates over three summer months (June-August). The top and bottom rows of Fig.\ref{fig:RPF} show the RPF of passive and active DG and SDG prosumers. We limit the daily time window to 8-18, as RPF are zero outside this range. The DG prosumer's RPF in the top row show how active consumption alone can reduce the RPF levels, especially at lower export rates (compare top left to top right panels). This is of an added benefit to the DSO as the RPF levels, under active consumption, become a function of export rates, which can be controlled by DSOs. The left panel in the bottom row shows how installing and optimally operating a battery, even under passive consumption, further reduced RPF (compare top left to bottom left panels). A more wiped-out (second row's middle and right panels) map is exhibited by active SDG prosumers, who maintained the lowest RPF levels, due to both storage and consumption acting to absorb PV output before it is exported to the grid. 

\subsection{Utility avoided costs under optimal co-optimization policy}
It is known that BTM generation helps utilities in avoiding costs related to generation, transmission and distribution systems. Avoided cost calculators (ACC) have been developed by many PUCs to quantify the {\em utility benefits} from BTM generation, covering various cost components including energy, transmission and distribution capacities and losses \cite{E3ACC}. As defined in \cite{Alahmed&Tong:22TSG}, the {\em utility net cost}\footnote{The utility net costs can be narrowly looked at as cost-shifts (cross-subsidies) if the utility directly transfers this cost to non-adopters.} ($\Psi$) is the difference between the customer {\em bill savings} $\Delta P^{\mbox{\tiny NEM}}_t(\cdot)$ due to BTM DER and the utility benefit (i.e., avoided cost). 
\begin{equation}\label{eq:UtilityNetCost}
    \Psi_t := \Delta P^{\mbox{\tiny NEM}}_t(g_t) - \pi^{\mbox{\tiny AC}}_t g_t,
\end{equation}
where $\pi^{\mbox{\tiny AC}}_t$ is the avoided cost rate determined by the ACC.

\par
\begin{figure}
    \centering
    \includegraphics[scale=0.43]{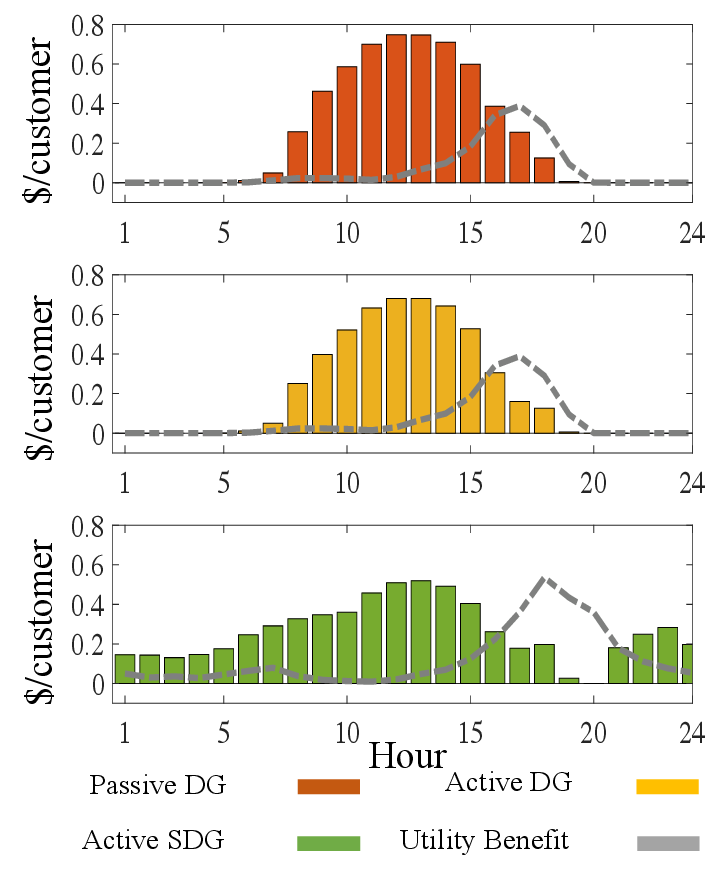}
    \vspace{-0.4cm}
    \caption{Utility net cost ($\bar{e}=\underline{e}=1$kW, $\pi^-=0.5\pi^+$).}
    \label{fig:NetCost}
\end{figure}

 Using California PUC's adopted ACC \cite{E3ACC}, we calculate the utility benefit of passive DG, active DG and active SDG prosumers in Fig.\ref{fig:NetCost}. For all customer types, the net cost is the highest during PV hours, because the utility benefit from BTM generation (gray curve) is small ($\pi^{\mbox{\tiny AC}}$ is small). Unlike SDG customers (bottom panel), DG customers (top and middle panels) created higher net costs to utilities during PV generation hours (8-16). Active DG prosumers (middle panel) incurred smaller net costs compared to passive ones (top panel) because more renewable generation is locally consumed creating fewer bill savings. The utility net costs during PV hours are the smallest under active SDG prosumers, but the net costs during non-PV hours increased due to battery discharging, indicating that the battery did not curtail, but rather shifted the utility net costs to no-PV hours \cite{Borenstein_canNetMetering}.

